\documentclass[twocolumn]{pasj00}

\Received{2015 February 20}
\Accepted{2015 May 8}
\SetRunningHead{N. Sakai et al.}{Astrometry of IRAS 07427-2400}

\usepackage{times}
\usepackage{supertabular}
\usepackage{graphicx}
\usepackage{lscape}     
\usepackage{cases}      
\usepackage{multicol}   
\usepackage[a4paper]{geometry}   
\usepackage{ulem}

\begin{document}

\title{Outer Rotation Curve of the Galaxy with VERA III: Astrometry of IRAS 07427-2400 and Test of the Density-Wave Theory}
\author{Nobuyuki Sakai$^{1}$, Hiroyuki Nakanishi$^{2}$, Mitsuhiro Matsuo$^{2}$, Nagito Koide $^{2}$, Daisuke Tezuka$^{2}$, Tomoharu Kurayama$^{3}$, Katsunori M. Shibata$^{1,4}$, Yuji Ueno$^{1}$, and Mareki Honma$^{1,4}$%
\thanks{Last update: January 19, 2007}}
\affil{%
$^{1}$Mizusawa VLBI Observatory, National Astronomical Observatory of Japan, Mitaka, Tokyo 181-8588}
\affil{%
$^{2}$Faculty of Science, Kagoshima University, 1-21-35 Korimoto, Kagoshima, Kagoshima, 890-0065}
\affil{%
$^{3}$Center for Fundamental Education, Teikyo University of Science, 2525 Yatsusawa, Uenohara, Yamanashi 409-0193}
\affil{%
   $^{4}$The Graduate University for Advanced Studies (Sokendai), Mitaka, Tokyo 181-8588}
\email{nobuyuki.sakai@nao.ac.jp}

\maketitle

\begin{abstract}
We report the trigonometric parallax of IRAS 07427-2400 with VERA to be 0.185 $\pm$ 0.027 mas, correspond.ing to a distance of 5.41$^{+0.92}_{-0.69}$ kpc. The result is consistent with the previous result of 5.32$^{+0.49}_{-0.42}$ kpc obtained by Choi et al. (2014) within error. To remove the effect of internal maser motions (e.g., random motions), we observed six maser features associated with IRAS 07427-2400 and determined systematic proper motions of the source by averaging proper motions of the six maser features. The obtained proper motions are ($\mu_{\alpha}$cos$\delta$, $\mu_{\delta}$) = ($-$1.79 $\pm$ 0.32, 2.60 $\pm$ 0.17) mas yr$^{-1}$ in equatorial coordinates, while Choi et al. (2014) showed ($\mu_{\alpha}$cos$\delta$, $\mu_{\delta}$) = ($-$2.43 $\pm$ 0.02, 2.49 $\pm$ 0.09) mas yr$^{-1}$ with one maser feature. Our astrometry results place the source in the Perseus arm, the nearest main arm in the Milky Way. Using our result with previous astrometry results obtained from observations of the Perseus arm, we conducted direct (quantitative) comparisons between 27 astrometry results and an analytic gas dynamics model based on the density-wave theory and obtained two results. First is the pitch angle of the Perseus arm determined by VLBI astrometry, 11.1 $\pm$ 1.4 deg, differing from what is determined by the spiral potential model (probably traced by stars), $\sim$ 20 deg. The second is an offset between a dense gas region and the bottom of the spiral potential model. The dense gas region traced by VLBI astrometry is located downstream of the spiral potential model, which was previously confirmed in the nearby grand-design spiral galaxy M51 in Egusa et al. (2011).   
\end{abstract}

\section{Introduction}
In 1964, Lin $\&$ Shu (1964) proposed the density-wave theory as a way of overcoming the winding dilemma, in which a spiral arm in a disk galaxy is destroyed within a time scale of several galactic rotations due to the differential rotation. Since then the density-wave theory has been used to discuss the evolution and dynamics of a spiral arm in a disk galaxy. For instance, Fujimoto (1968) and Roberts (1969) examined non-linear gas motions perturbed by the density wave (spiral arm) numerically, and found the ``Galactic shock'' where a velocity jump (non-circular motion) occurred and the gas accumulated upstream of the spiral potential (see fig. 5 in Roberts 1969). Burton (1973) and Mel'nik et al. (1999) schematically showed the systematic non-circular motion based on the density-wave theory.

In contrast to the density-wave theory, another dynamics model, the recurrent transient spiral, has been proposed based on N-body simulations with or without a gas disk (e.g., Miller et al. 1970; Hohl 1971; Hockney $\&$ Brownrigg 1974; James $\&$ Sellwood 1978; Sellwood $\&$ Carlberg 1984; Baba et al. 2010; Wada et al. 2011; Fujii et al. 2011; Grand et al. 2012; Baba et al. 2013). In the recurrent transient spiral, the spiral arm is destroyed and regenerated over several rotational periods. Wada et al. (2011) reported that gas flows converge near the bottom of the spiral potential with random motions, which is different from the result based on the density-wave theory.

Ever since the 1960s, the two models have been proposed and suggested by simulations and observations. However, the previous observations were mainly based on line-of-sight velocities (1D information) and apparent 2D positions on the celestial sphere, which were not enough to conduct quantitative comparisons between the observations and the simulations in terms of both velocity and spatial resolutions. 

Since the 2000s the Very Long Baseline Interferometry (VLBI) technique has allowed us to routinely conduct Galactic-scale astrometry and to obtain accurate information about 3D positions and velocities of Galactic masers in the Milky Way (e.g., Reid $\&$ Honma 2014a; Reid et al. 2014b). At the moment, we can conduct direct (quantitative) comparisons between the astrometry results and the dynamical models for understanding the dynamics and evolution of the spiral arm, as well as those of the Milky Way.

VLBI astrometry observations have revealed systematic peculiar (non-circular) motions in the Perseus arm (e.g., Sakai et al. 2012, 2013; Choi et al. 2014). The Perseus arm traced by stars and gas is one of the Milky Way's main arms (e.g., Churchwell et al. 2009). Sakai et al. (2012) and (2013) studied the 3D structure and kinematics of the Perseus arm using VLBI astrometry results. The sources in the Perseus arm were moving systematically toward the Galactic Center and lagged behind the Galactic rotation. 

Sakai et al. (2012) determined averaged peculiar motions of ($U_{{\rm mean}}$, $V_{{\rm mean}}$) = (11 $\pm$ 3, $-$17 $\pm$ 3) km s$^{-1}$ with seven sources in the Perseus arm with $\sim$ 4-$\sigma$ significance. Note that $U$ and $V$ are directed to the Galactic center and the Galactic rotation, respectively. The systematic peculiar motions were also confirmed by Choi et al. (2014), who determined averaged peculiar motions of ($U_{{\rm mean}}$, $V_{{\rm mean}}$) = (9.2 $\pm$ 1.2, $-$8.0 $\pm$ 1.3) km s$^{-1}$ with 25 sources in the Perseus arm. The systematic peculiar motions are consistent with the Galactic shock proposed by Fujimoto (1968) and Roberts (1969). Thus, the Perseus arm is a good target for testing the density-wave theory.

IRAS 07427-2400, a high-mass star-forming region (Qiu et al. 2009), had the largest Galactic longitude ($l$ $\sim$ 240$^{\circ}$) in the Perseus arm in Choi et al. (2014). The source played an important role in the study of the large-scale structure and kinematics of the Perseus arm (e.g., determination of the pitch angle). However, Choi et al. (2014) observed only one maser feature to discuss the systemic motion of the source (e.g., Galactic rotation), while Qiu et al. (2009) confirmed the bipolar molecular outflow of $^{13}$CO(J=2-1) around the source. To discuss the systemic motion of the source precisely, more maser features should be observed to remove the effect of internal maser motions (e.g., outflow motions). 

In addition, a different distance has rarely been measured for the same source with independent VLBI arrays (e.g., $d$ = 5.03 $\pm$ 0.19 kpc for G048.60+0.02 in Nagayama et al. 2011; $d$ = 10.75$^{+0.61}_{-0.55}$ kpc for G048.60+0.02 in Zhang et al. 2013), although the same distances have been measured for the same sources in most cases. Therefore, independent distance measurements with independent arrays are important to confirm astrometry results precisely.
    
In this paper, we will report astrometry results of IRAS 07427-2400 with VERA for discussing the large-scale structure and kinematics of the Perseus arm more precisely. Using our results with previous astrometry results, we will compare the results with an analytic gas dynamics model for testing the density-wave theory.

\begin{table*}[t]
\begin{center}
\small
\caption{Observations.}
\begin{tabular}{lccccc}
\hline
\hline
Epoch	&Date	&Time Range	&Beam	&Image r.m.s. &Detected maser feature\\ 
  		&	&(UTC)		&(mas)	&(mJy/beam)		\\
\hline
  A	&Jan 18, 2012&11:00-18:05 &1.7$\times$0.8@157$^{\circ}$  &124	&1, 2, 3, 4\\
  B	&Feb 28, 2012&08:20-15:25 &1.7$\times$0.8@157$^{\circ}$  &180	&1, 2, 3, 4\\
  C	&Apr 24, 2012&04:40-11:45 &1.7$\times$0.8@162$^{\circ}$  &154	&1, 2, 3, 4, 5\\
  D	&Jun 17, 2012&01:10-08:15 &1.7$\times$0.7@157$^{\circ}$  &356	&\\
  E	&Sep 10, 2012&19:30-02:35 &1.9$\times$0.9@162$^{\circ}$  &384	&\\
  F	&Jan 27, 2013&09:55-18:20 &1.9$\times$0.8@162$^{\circ}$  &69		&1, 2, 3, 4, 5, 6\\
  G	&Mar 20, 2013&06:30-14:55 &1.8$\times$0.9@158$^{\circ}$  &159	&1, 2, 3, 4, 5, 6\\
  H	&Jun 02, 2013&01:35-10:00 &1.9$\times$0.8@159$^{\circ}$  &249	&1, 2, 3, 4, 5, 6\\
  I	&Sep 16, 2013&18:35-03:00 &1.9$\times$0.8@160$^{\circ}$ 	&103 	&1, 2, 3, 4, 5, 6\\

\hline
\multicolumn{0}{@{}l@{}}{\hbox to 0pt{\parbox{180mm}{\footnotesize
\par\noindent
}\hss}}
\end{tabular}
\end{center}
\end{table*}

\begin{table*}[th]
\caption{Source Data.}
\begin{tabular}{lccccc}
\hline
\hline
Source Name	&R.A.	&Decl.	&S.A.\footnotemark[$*$]	&Flux Density	&Note	\\
		&(J2000.0)	&(J2000.0)				&($^\circ$)			&(Jy)	&\\
\hline
IRAS 07427-2400	&\timeform{07h44m51.9205s}	&\timeform{-24d07'41".457}	&	&11 $\sim$ 27	&H$_{2}$O masers	\\
J0745-2451		&\quad \timeform{07h45m10.263445s}\footnotemark[$\dag$]        &\timeform{-24d51'43".76681}\footnotemark[$\dag$]	&0.74	&0.07 $\sim$ 0.24	&Phase-reference source	\\
\hline
\multicolumn{4}{@{}l@{}}{\hbox to 0pt{\parbox{180mm}{\footnotesize
\par\noindent
\footnotemark[$*$]Separation angle between the maser and reference sources.\\
\footnotemark[$\dag$]The positions are based on Petrov et al. (2011).
}\hss}}
\end{tabular}
\end{table*}

\section{Observations and Data Reduction}
\subsection{VLBI Observations with VERA}
We carried out VERA astrometry observations of H$_{2}$O maser at a rest frequency of 22.235080 GHz to measure the trigonometric parallax and proper motions of IRAS 07427-2400. Note that IRAS 07427-2400 was observed as a part of the Outer Rotation Curve project with VERA (as explained in Sakai et al. 2012), aiming to understand the mass distribution of the Milky Way especially for the outer disk based on an accurate rotation curve. The observations procedure has been described in previous VERA astrometry papers (e.g., Sakai et al. 2012). 

In table 1, the observation periods and synthesized beam sizes are listed with image noise levels and information about detected maser features. In table 2, maser and reference source information are listed for phase-referencing observations (e.g., tracking-center positions, separation angle between the two sources, and flux densities). Note that we observed one or two bright continuum source(s) in each observation to conduct a clock offset calibration of each VERA station. The observed continuum sources were 3C84, DA193, M87, OJ287, and BLLAC from which we selected one or two source(s) in each observation.

As for an on-source time of the maser source in each observation, $\sim$ 2 hours were assigned in the total observation time of $\sim$ 7 hours in the five former observations, while $\sim$ 3 hours were assigned in the total observation time of $\sim$ 8 hours in the four later observations (as shown in table 1). Note that we observed two pairs of maser and reference sources for VLBI astrometry observations. One of them is listed in table 2 and reported in this paper. The other pair, IRAS 07207-2400 and J0729-1320, will be reported in another paper. For velocity resolution of the maser source we set 0.42 km s$^{-1}$ with a velocity coverage of $\sim$ 215 km s$^{-1}$ , instead of a high resolution mode of 0.21 km s$^{-1}$ with a velocity coverage of $\sim$ 108 km s$^{-1}$ (e.g., explained in Sakai et al. 2014).

\subsection{Data Reduction}
We conducted general phase-referencing analysis with AIPS (Astronomical Image Processing System, NRAO) in the same manner as Kurayama et al. (2011). The main analysis procedures are described as follows (see more detail in the Appendix of Kurayama et al. 2011):
\begin{itemize}
\item[1.] Amplitude calibration for line (maser) and continuum (phase-reference and clock-calibration) sources
\item[2.] Accurate recalculation of a tracking model used in the Mitaka FX correlator
\item[3.] Binding of frequency channels for the continuum sources
\item[4.] Calibration of clock parameters with the clock calibrator
\item[5.] Calibration of clock parameters with the reference source
\item[6.] Imaging of the reference source after the self calibration
\item[7.] Phase referencing for the maser source (subtracting phases of the reference source from those of the maser source)
\item[8.] Dual-beam calibration with the ``horn-on-dish'' method (Honma et al. 2008)
\item[9.] Calibration of the Doppler effect for the maser source
\item[10.] Making a CLEANed image of the maser source with the phase referencing in each velocity channel
\item[11.] Determining an absolute position and a flux density of the maser source in each velocity channel.
\end{itemize}
Step 2 (accurate recalculation of a tracking model) was needed, since the tracking model used in the correlator did not have sufficient accuracy for astrometry. For step 11 (determining an absolute position and a flux density of the maser), we conducted an elliptical Gaussians fit with the task ``jmfit'' in AIPS. 
 
 We regarded a maser as detected if it achieved a high SNR (Signal-to-Noise Ratio) of five or more. We traced a maser detected in the same velocity channel and its position movements during the observations to determine the parallax and proper motions. We determined the proper motions using masers detected with three epochs or more. On the other hand, we determined the parallaxes using masers detected in at least two continuous velocity-channels with five epochs or more.

On determinations of the parallax and proper motions, we used the ``VERA Parallax'', one of the tasks of ``VEDA (VEra Data Analyzer)'', a data analyzing software developed at NAOJ. In the task, we assumed that source motions can be modeled by a combination of linear proper and sinusoidal parallax motions.

\begin{figure}[tb]
 \begin{flushleft} 
    \includegraphics[width=70mm, height=70mm]{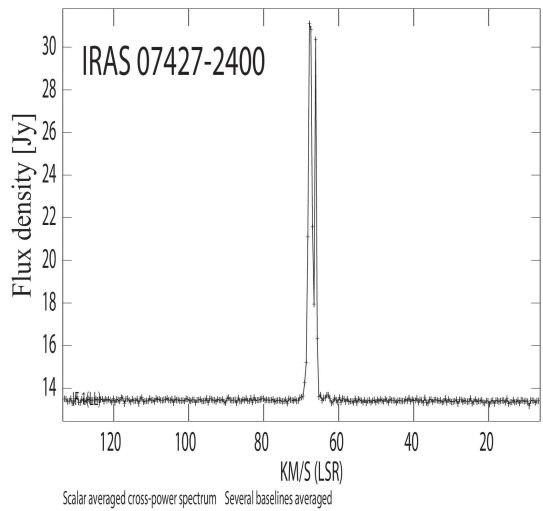}
\end{flushleft}
\caption{ Scalar-averaged cross-power spectrum of IRAS 07427-2400 taken at epoch F with several baselines averaged. }
\label{fig1}
\end{figure}

\begin{table*}
\begin{center}
\caption{Parallax fits.\footnotemark[$*$]}
\small
\begin{tabular}{cccccccc}
\hline
\hline
Feature	&$V_{\rm{LSR}}$	&N$_{epochs}$	&Epochs	&Parallax(Error)	&	&\multicolumn{2}{c}{Errors	}\\
 \cline{7-8}
		&km s$^{-1}$	&				&			&(mas)	&	&R.A.	&Dec.	\\
		&	&				&			&	&	&\multicolumn{2}{c}{(mas)}	\\
\hline
1	&68.7	&6	&ABC++FG+I		&0.171(0.039)		&	&0.049		&0.162		\\
	&68.2	&7	&ABC++FGHI		&0.162(0.047)		&	&0.066		&0.200		\\
	&67.8	&7	&ABC++FGHI		&0.184(0.064)		&	&0.094		&0.171		\\
	&67.4	&7	&ABC++FGHI		&0.182(0.073)		&	&0.110		&0.180		\\
	&67.0	&6	&A+C++FGHI		&0.181(0.075)		&	&0.109		&0.195		\\


2	&66.6	&7	&ABC++FGHI		&0.193(0.029)		&	&0.041		&0.230		\\
	&66.1	&7	&ABC++FGHI		&0.205(0.020)		&	&0.028		&0.201		\\
	&65.7	&6	&A+C++FGHI		&0.163(0.035)		&	&0.048		&0.217		\\

3	&64.9	&7	&ABC++FGHI		&0.225(0.018)		&	&0.025		&0.296		\\
	&64.4	&6	&ABC++FG+I		&0.218(0.030)		&	&0.037		&0.132		\\

\hline
\multicolumn{2}{l}{Combined fit for 10 spots}					&	&	&0.185(0.015)	&		&0.069	&0.204		\\
\hline
\multicolumn{3}{l}{Final}		&	&0.185(0.027)	&		&	\\

\hline

\multicolumn{4}{@{}l@{}}{\hbox to 0pt{\parbox{130mm}{\footnotesize
\par\noindent
\\
\footnotemark[$*$]A combined fit was done to a data set of 10 spots (see text). The error of the final parallax was estimated to multiply the error of the combined fit by $\sqrt{\frac{N_{\rm{spot}}}{N_{\rm{feature}}}}$ where $N_{\rm{spot}}$ and $N_{\rm{feature}}$ are the numbers of maser spots and features, respectively (see also text).
}\hss}}
\end{tabular}
\end{center}
\end{table*}

\begin{table*}[tbp]
\begin{center}
\caption{Determination of systematic proper motions for IRAS 07427-2400.}
\small
\begin{tabular}{ccccc}
\hline
\hline
Feature	&	&\multicolumn{2}{c}{Proper Motions\footnotemark[$\ast$] (Error)}	&Note\\
\cline{3-4}
	&$V_{\rm{LSR}}$	&$\mu_{\alpha} \rm{cos}\delta$	&$\mu_{\delta}$	&\\
	&km s$^{-1}$	&(mas yr$^{-1}$)&(mas yr$^{-1}$)	&\\
\hline
1 			&69.9 $\sim$ 66.6	&$-$2.55	&2.51	&		\\
2			&66.6 $\sim$ 65.7 	&$-$1.45	&2.10	&		\\
3			&64.9 $\sim$ 64.4 	&$-$0.53	&3.31	&		\\
4			&67.0 $\sim$ 66.6	&$-$2.44	&2.83	&		\\
5			&70.4 $\sim$ 68.7 	&$-$1.50	&2.32	&		\\
6			&63.6 $\sim$ 62.3 	&$-$2.26	&2.51	&		\\

\hline
Mean		&66.4\footnotemark[$\dag$]				 	&$-$1.79 (0.32)\footnotemark[$\ddag$]	&2.60 (0.17)\footnotemark[$\ddag$]	&Systematic proper motions		\\
\hline
\multicolumn{4}{@{}l@{}}{\hbox to 0pt{\parbox{180mm}{\footnotesize
\par\noindent \\
\footnotemark[$\ast$]
Proper motions were determined by adapting a parallax of 0.185 mas.  \\
\footnotemark[$\dag$]
The value was determined by averaging the results of the six features.  \\
\footnotemark[$\ddag$]
The standard errors.  \\

}\hss}}

\end{tabular}
\end{center}
\end{table*}

\begin{figure*}[tbp]
 \begin{center}
\includegraphics[width=159mm, height=157mm]{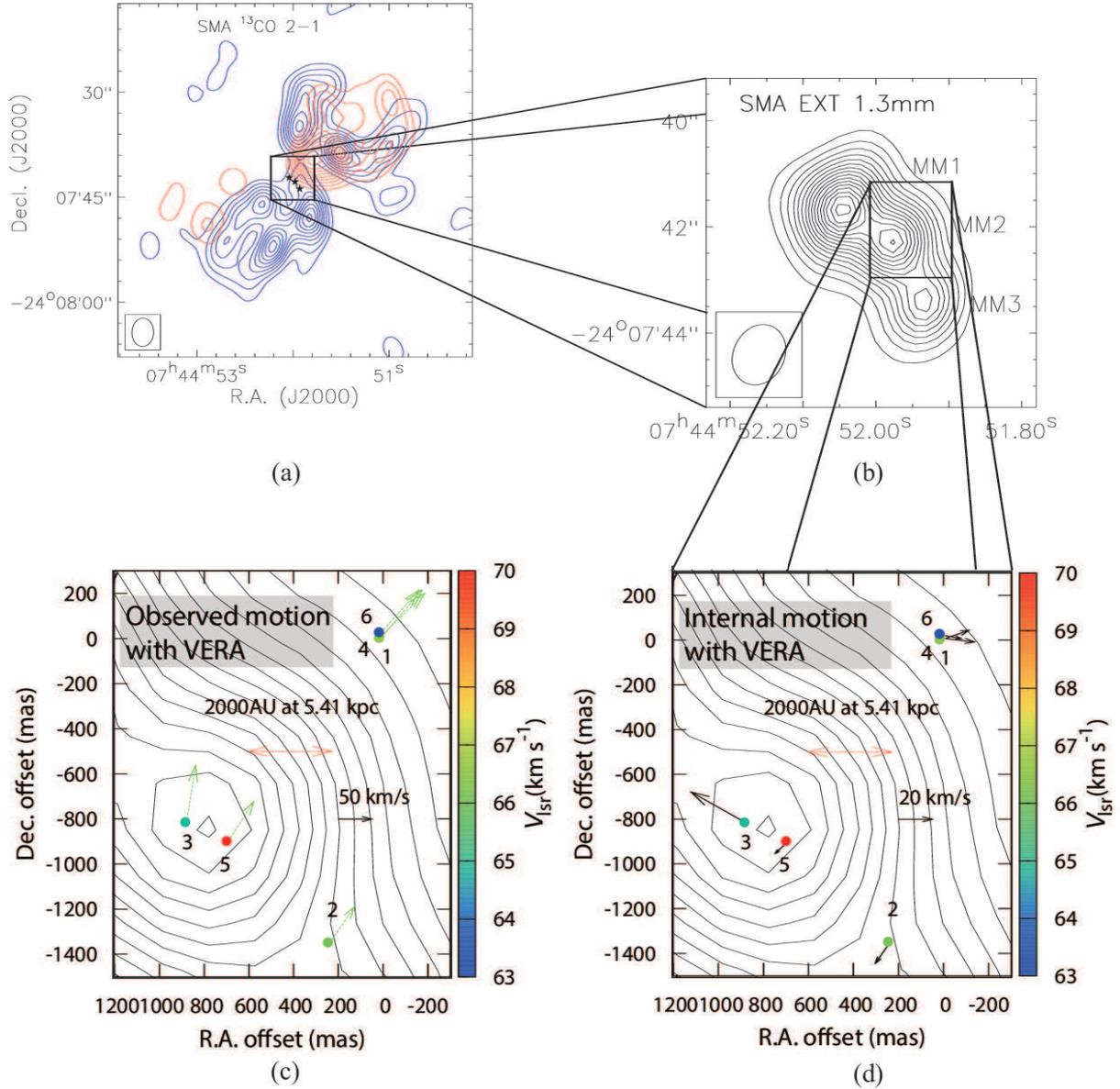}
\end{center}
\caption{ (a) Large-scale view of IRAS 07427-2400 with which the bipolar outflow of $^{13}$CO(J=2-1) is associated (Qiu et al. 2009). Red and blue contours represent redshifted and blueshifted $^{13}$CO(J=2-1) lobes, respectively. Redshifted and blueshifted lobes were integrated from 74 to 78 km s$^{-1}$ and from 58 to 62 km s$^{-1}$, respectively. (b) Magnified view of the rectangle of fig. 2a. Contours show continuum (1.3 mm) emissions in which there are three 1.3-mm peaks as MM1, MM2, and MM3 (Qiu et al. 2009). (c) Maser spatial and velocity distributions (as colored circles) with observed motions (as green arrows) relative to the phase reference source in table 2. The distributions were made based on epoch F in table 1. We adopted our distance measurement of 5.41 kpc to show the observed motions in absolute velocities. The numbers labeled correspond to the maser features listed in table 1. Nominal origin is set to the maser detection position in table 2. (d) Same as (c), but with systematic proper motions subtracted from fig. 2(c) (i.e., internal motions of masers) (see text).}
\label{fig2}
\end{figure*}


\section{Results}
\subsection{Maser spatial/velocity distributions in IRAS 07427-2400}
During the observations over a period of one year and eight months, we detected maser emissions except for epochs D and E as shown in table 1. For epochs D and E we could not make good quality images due to high image noise levels (e.g., table 1). Figure 1 shows a typical maser spectrum taken at epoch F. 

Figure 2a shows a large-scale view of IRAS 07427-2400 with which the bipolar molecular outflow of $^{13}$CO(J=2-1) is associated (Qiu et al. 2009). Figure 2b represents the magnified view of fig. 2a, and there are three continuum (1.3 mm) peaks as MM1, MM2, and MM3 (Qiu et al. 2009) in fig. 2b. Figures 2c and 2d show VERA observation results in the same area of the rectangular area in fig. 2b. Labeled numbers in figures 2c and 2d represent maser features listed in table 1. The maser feature is recognized as a maser cluster, which consists of closest maser spots in continuous velocity channels. In figure 2c, one can see maser spatial/velocity distributions and observed motions relative to the reference source listed in table 2.

\subsection{Trigonometric parallax of IRAS 07427-2400}
As a result of the data analyses, we detected three maser features to determine the parallaxes as shown in table 3. Using the three maser features consisting of 10 spots, we determined the parallax of IRAS 07427-2400 to be 0.185$ \pm$ 0.015 mas in table 3. The position errors shown in table 3 for right ascension (R.A.) and declination (Decl.) were introduced so that the reduced $\chi^{2}$ became unity, since systematic error is the dominant source of error in VLBI astrometry observation (e.g., Sanna et al. 2012). The position error of R.A., 0.069 mas, is smaller than that of Decl., 0.204 mas, which is consistent with previous VERA results. As for the obtained parallax error (0.015 mas), it may be underestimated since the 10 spots may not be independent of each other. For a conservative parallax estimation, we multiplied the obtained error by $\sqrt{\frac{N_{\rm{spot}}}{N_{\rm{feature}}}}$ where $N_{\rm{spot}}$ and $N_{\rm{feature}}$ are the numbers of maser spots ($N_{\rm{spot}}$ = 10) and features ($N_{\rm{feature}}$ = 3), respectively.  

\begin{figure*}[tbp]
\includegraphics[width=80mm, height=80mm]{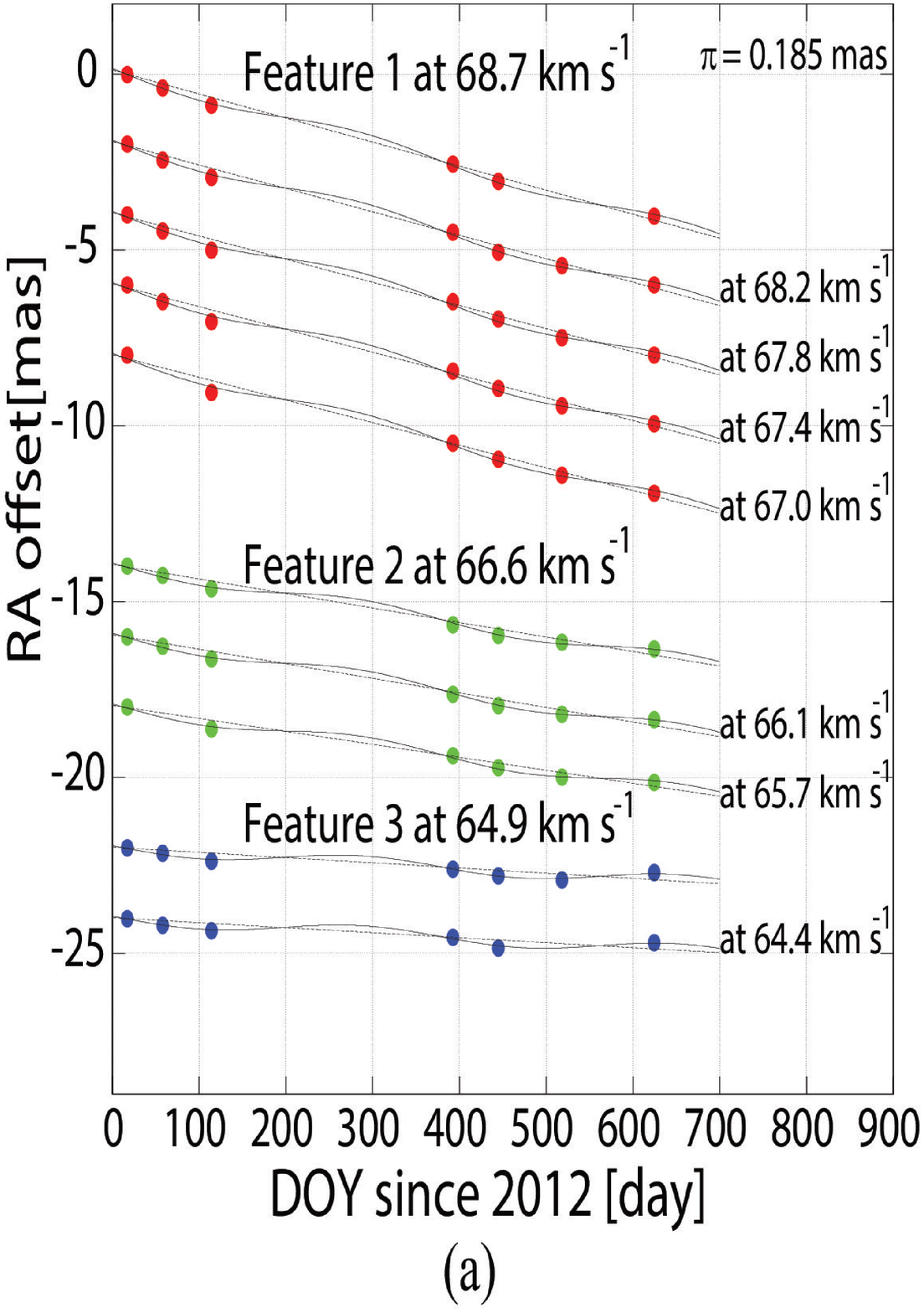}%
\includegraphics[width=80mm, height=80mm]{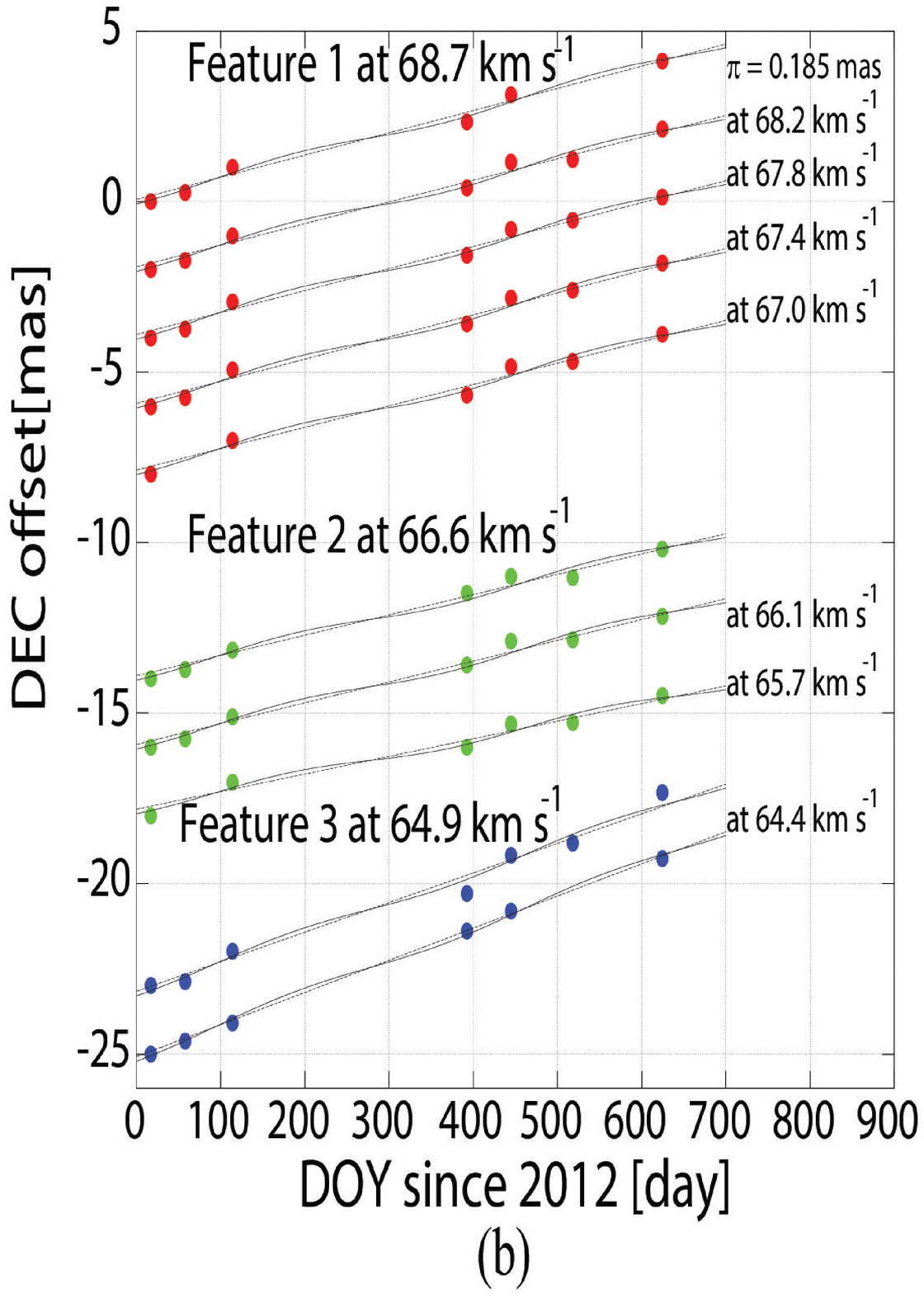}\\
\includegraphics[width=80mm, height=60mm]{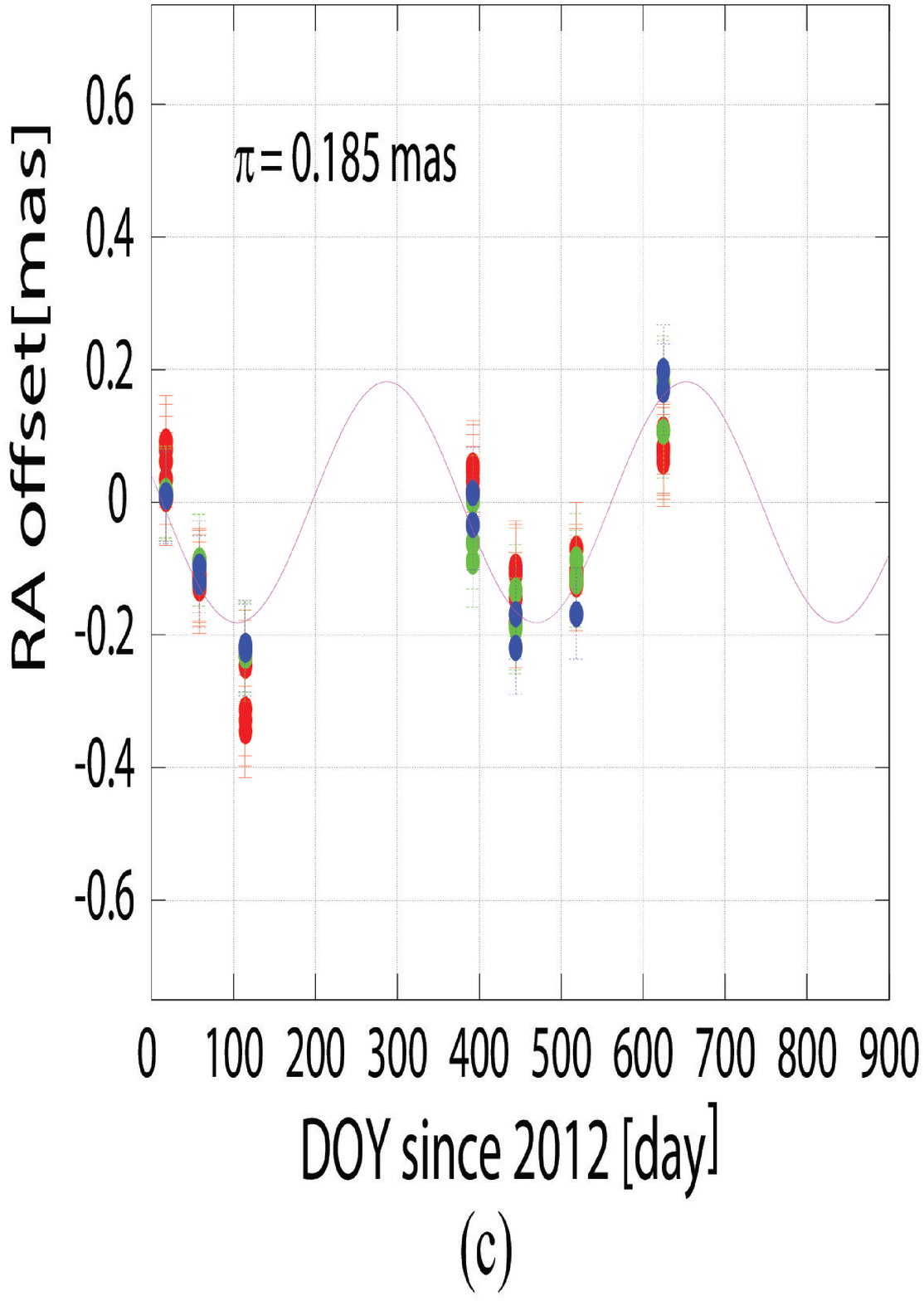}%
\includegraphics[width=80mm, height=60mm]{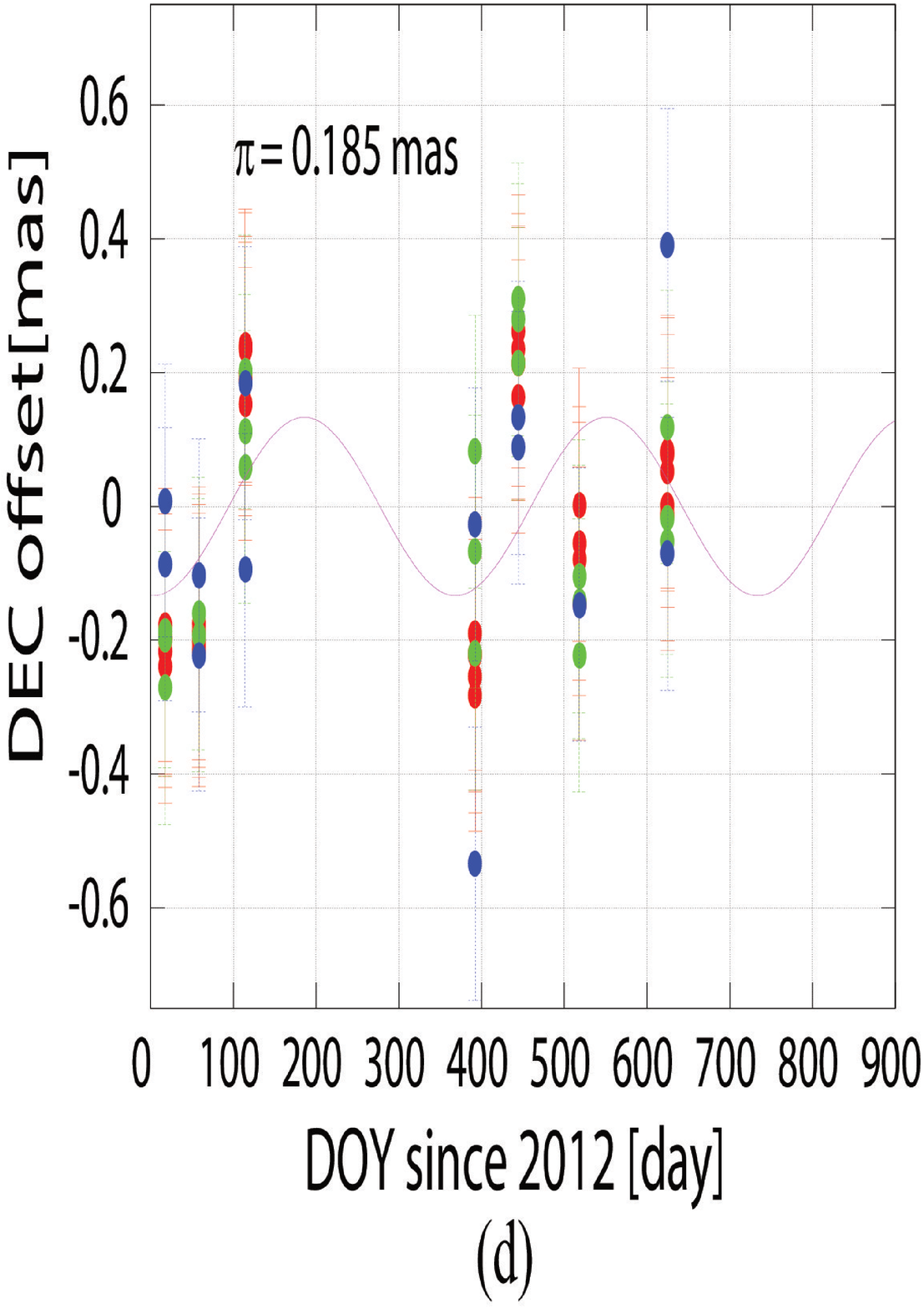}\\
\caption{Maser positional evolutions and combined-fit results for the 10 maser spots listed in table 3 (see text). Error bars represent position errors resulting from systematic errors, which are introduced so that the reduced $\chi^{2}$ becomes unity. Different colors correspond to respective features.
(a) The maser's positional evolutions in right ascension as shown by circles. Solid lines show the parallax and proper motion fits (see text), while dashed lines represent only the proper motion fits. (b) Same as (a), but in declination.
 (c) Parallax motions (as the circles) in right ascension after subtracting the proper motions from (a). The solid curve shows the modeled parallax motion. (d) Same as (c), but in declination.
}
\end{figure*}

We obtained the final parallax to be 0.185 $\pm$ 0.027 mas, corresponding to a distance of 5.41$^{+0.92}_{-0.69}$ kpc. The distance is consistent with the previous parallactic distance of 5.32$^{+0.49}_{-0.42}$ kpc determined by Choi et al. (2014) within error. On the other hand, the distance is smaller than the previously estimated source distance of 6.9 kpc, based on the kinematic distance derived from $^{12}$CO(J=1-0) observations (Wouterloot $\&$ Brand 1989). The difference between the distances can be explained using peculiar (non-circular) motions of the source. We will further discuss the peculiar motions in chapter 4.

Figure 3 shows combined fit results in which maser motions are modeled by a combination of common sinusoidal parallax and discrete linear proper motions. Figures 3a and 3b show the combined fit results of each maser spot with the final parallax result, fixed in directions of R.A. and Decl., respectively. Error bars in each panel of figure 3 represent the systematic position errors explained previously. Figures 3c and 3d show the sinusoidal parallax motions after subtracting the proper motions from figures 3a and 3b in directions of R.A. and Decl., respectively. Large deviations from the model are seen in Decl. (figure 3d), which have been shown in previous VLBI astrometry results (e.g., Sakai et al. 2012). The main reason for the deviation is thought to be tropospheric zenith delay residuals as discussed in Honma et al. (2007).

\subsection{Systematic proper motions of IRAS 07427-2400}
Since observed proper motions include systemic (e.g., Galactic rotation) and internal maser (e.g., outflow) motions, we should separate both motions for discussing the systemic motion of IRAS 07427-2400. Table 4 displays observed proper motions of several maser features that were detected with three epochs or more. Note that the proper motion components were determined by adapting a common parallax of 0.185 mas.   

In fig. 2c we can see clear northwest motions, which represent the Galactic rotation. The velocity range of maser features in fig. 2c is not consistent with velocity ranges of redshifted and blueshifted lobes of $^{13}$CO molecular outflow (Qiu et al. 2009). Thus, we assumed that internal maser motions can be regarded as random motions, which means that we can determine the systematic proper motions by averaging proper motions of several maser features. Averaging of multiple features is essential for obtaining the systematic proper motions, since table 4 shows that a difference of observed proper motions is up to $\sim$ 2 mas, corresponding to $\sim$ 50 km s$^{-1}$ at a distance of 5.4 kpc. The difference is caused by internal maser motions.

As a result of the data averaging in table 4, we determined the systematic proper motions to be ($\mu_{\alpha}$cos$\delta$, $\mu_{\delta}$) = ($-$1.79 $\pm$ 0.32, 2.60 $\pm$ 0.17) mas yr$^{-1}$ in equatorial coordinates. Note that the obtained errors were determined by dividing standard deviations by a factor of \ $\sqrt[]{\rm{n}}$, where n is the number of measurements (e.g., n = 6 applied in this case as shown in table 4). Table 4 shows a mean velocity ($V_{{\rm LSR}}$) of $\sim$ 66 km s$^{-1}$, which is $\sim$ 2 km s$^{-1}$ smaller than a velocity of CS(J=2-1) emission in IRAS 07427-2400 (Bronfmann et al. 1996). The difference indicates that the systematic proper motions include errors of at least $\sim$ 2 km s$^{-1}$. In fact, the obtained errors in table 4, 0.32 mas and 0.17 mas, are converted to $\sim$ 8 km s$^{-1}$ and $\sim$ 4 km s$^{-1}$ at a distance of 5.4 kpc, respectively.

 Figure 2d represents internal maser motions after subtracting the systematic proper motions from fig. 2c, which in fact seems to be the random motions. We will discuss the systemic motion of IRAS 07427-2400 using the resultant systematic proper motions in chapter 4.

\subsection{Peculiar motions of IRAS 07427-2400}
As the next step, we determine the peculiar (non-circular) motions of IRAS 07427-2400. Full-space (3D) motions of the source can be determined from the observed 3D position, systematic proper motions, and systemic velocity of the source. For the systemic velocity we refer to 68.0 $\pm$ 5.0 km s$^{-1}$ obtained from CS(J=2-1) observations (Bronfmann et al. 1996). Calculation procedures for determining the 3D motion are itemized as follows:
\begin{itemize}
\item[1.] Velocity conversion from the LSR velocity ($V_{\rm{LSR}}$) to the heliocentric velocity ($v_{\rm{helio}}$)
\item[2.] Coordinates conversion from equatorial coordinates to Galactic coordinates  
\item[3.] Conversion from the proper (apparent) motions to absolute motions using the parallax result ($\pi$)
\item[4.] Corrections of the peculiar solar motions, the Galactic constants ($R_{0}$ and $\Theta_{0}$), and the Galactic rotation curve ($\Theta$($R$))
\end{itemize}
\begin{figure*}[tb]
 \begin{center} 
     \includegraphics[width=157mm, height=84.0mm]{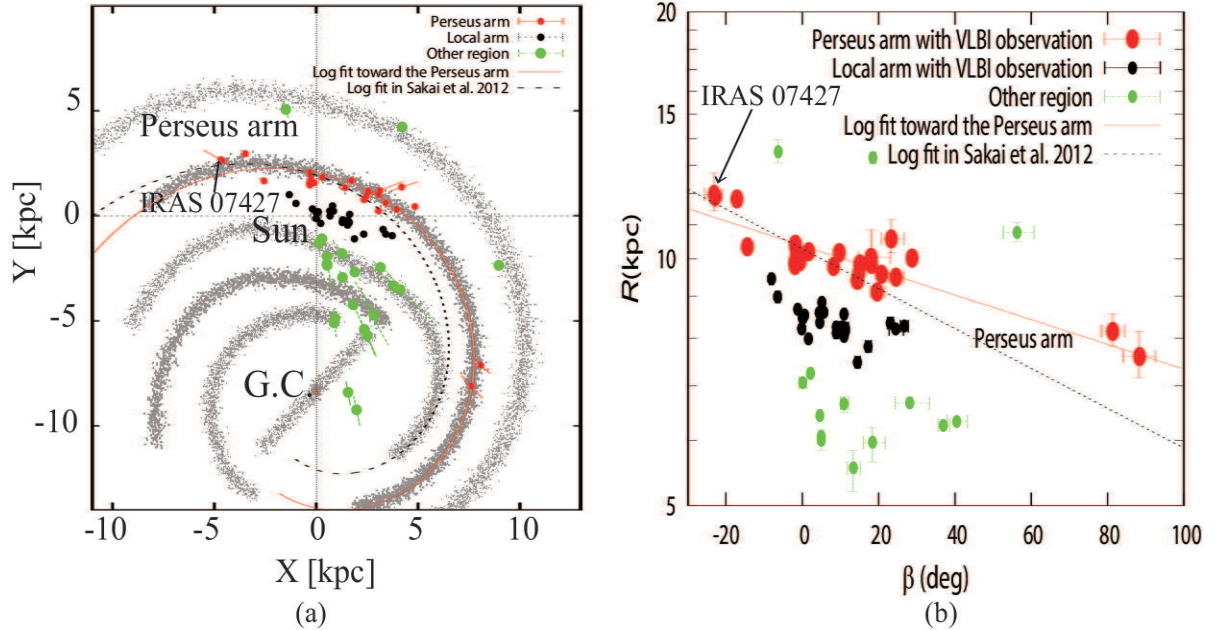}
\end{center}
\caption{\vtop{\hsize=400pt (a) Schematic face-on view of the Milky Way (Taylor $\&$ Cordes 1993 with updates) on which our astrometry result of IRAS 07427-2400 and previous astrometry results are superimposed. $R_{0}$ = 8.33 kpc (Gillessen et al. 2009) was assumed, and the Sun is located at the origin. Red and black circles represent the sources in the Perseus and Local arms, respectively, while green ones, referred from Sakai et al. (2012), show the sources in the other regions.  (b) Galactocentric azimuth ($\beta$) versus Galactocentric distance ($R$) diagram on which the astrometry results are superimposed. Note that $R$ axis is shown in the log scale. The solid line based on a logarithmic spiral model (see eq. 1 in text) was fitted to the sources in the Perseus arm, while the dashed line shows the previous result obtained by Sakai et al. (2012).}}
\label{fig4}
\end{figure*}

Note that detailed explanations for the above steps are summarized in the {\bf Appendix 1}. 

In Step 1 (Velocity conversion) $V_{{\rm LSR}}$ = 68.0 $\pm$ 5.0 km s$^{-1}$ is converted to $v_{\rm{helio}}$ = 86.4 $\pm$ 5.0 km s$^{-1}$ for the source. For Steps 2 and 3 we use the matrix multiplication shown in the {\bf Appendix 1}.  In Step 4 we adopt peculiar solar motions of ($U_{\odot}$, $V_{\odot}$, $W_{\odot}$) = (11.1 $\pm$ 1, 12.24 $\pm$ 2, 7.25 $\pm$ 0.5) km s$^{-1}$ (Sch\rm{\"o}nrich, Binney, $\&$ Dehnen 2010 ) and Galactic constants of ($R_{0}$, $\Theta_{0}$) = (8.33 kpc, 240 km s$^{-1}$) referred from Gillessen et al. (2009) and Reid $\&$ Brunthaler (2004). For the Galactic rotation curve we simply assume a flat rotation curve as $\Theta(R)$ = $\Theta_{0}$.

As a result, we obtain the peculiar motions of IRAS 07427-2400 to be ($U$, $V$, $W$) = ($-$8 $\pm$ 8, 0 $\pm$ 9, 1 $\pm$ 11) km s$^{-1}$. Directions of the peculiar motions are toward the Galactic center ($U$), the Galactic rotation ($V$), and the north Galactic pole ($W$). Note that here the obtained errors are estimated considering contributions from the parallax ($\delta_{\pi}$), the proper motions ($\delta_{\mu_{\alpha}\rm{cos}\delta}$, $\delta_{\mu_{\delta}}$), and the systemic velocity ($\delta_{v_{\rm{helio}}}$) in basically the same manner as Johnson $\&$ Soderblom (1987). A detailed explanation of the error calculation is described in the {\bf Appendix 1}.

\section{Discussion}

\subsection{Large-scale structure of the Perseus arm}
In this section, we determine a pitch angle and a reference position of the Perseus arm based on our parallax result and previous VLBI astrometry results. First, a combination of coordinates (e.g., Galactic longitude and latitude) and the parallax of IRAS 07427-2400 tells us the absolute source position on the Galactic plane as shown in fig. 4a. Note that we assumed a distance to the Galactic center (i.e., the Galactic constant $R_{0}$) of 8.33 kpc referred from Gillessen et al. (2009). The source is positioned only $\sim$ 6 pc away from the disk toward the north Galactic pole. Second, based on the source location and the longitude-velocity ($l$-$v$) diagram of CO(J=1-0) emission (Dame et al. 2001), the source is positioned in the Perseus arm (e.g., figures 4a and 4b, and fig. 3 in Dame et al. 2001). A shape of the spiral arm is often described with a logarithmic-spiral model (e.g., Reid et al. 2009a) as
\begin{equation}
\mathrm{ln}(R/R_{\rm{ref}})=-(\beta-\beta_{\rm{ref}})\mathrm{tan}i,
\end{equation}
where $i$ is the pitch angle of the spiral arm and $\beta$ is Galactocentric azimuth ($\beta$ is defined as 0 deg toward the Sun from the Galactic Center and increasing clockwise). ($R_{{\rm ref}}$, $\beta_{{\rm ref}}$) are reference positions of the spiral arm. Since ($R_{{\rm ref}}$, $\beta_{{\rm ref}}$) can have an infinity of combinations in the log-spiral fit, we regard $\beta_{{\rm ref}}$ as 0 deg in this paper for simplicity. Finally, we apply the log-spiral model to our result and the previous ones (as summarized in fig. 4b and table 5), then we can determine a pitch angle and a reference position of the Perseus arm. 

As a result, we obtain the parameters, ($i$, $R_{{\rm ref}}$), to be (11.2 $\pm$ 1.4 deg, 10.4 $\pm$ 0.1 kpc) using an unweighted least-squares fit toward 27 sources in the Perseus arm. If we exclude W49N and G48.60+0.02 located in the Perseus arm of the first Galactic quadrant from the fit, ($i$, $R_{{\rm ref}}$) = (12.2 $\pm$ 2.1 deg, 10.4 $\pm$ 0.1 kpc) are obtained, which are consistent with the results obtained for 27 sources within errors. The former pitch angle is also consistent with the previous result of 9.9 $\pm$ 1.5 deg determined by Choi et al. (2014) using 25 sources in the Perseus arm within error. 
 
On the other hand, the pitch angle (11.2 $\pm$ 1.4 deg) is not consistent with the previous result of 17.8 $\pm$ 1.7 deg determined by Sakai et al. (2012) using seven sources in the Perseus arm. The pitch angle difference (11.2 $\pm$ 1.4 deg or 17.8 $\pm$ 1.7 deg) may originate in the difference of the source numbers (27 or 7). However, here we cannot choose one as the best solution, since an observation gap is seen in the Perseus arm (e.g., 28.8 deg $\leq$ $\beta$ $\leq$ 81.4 deg in fig. 4b). More astrometry observations especially for the gap region would help us to judge whether a spur or bifurcation of the spiral arm, as seen in external disk galaxies, exists or not in the Perseus arm.

\newgeometry{left=1.8cm}
{\tabcolsep = 1.3mm
\begin{longtable}{llllllllll}
\caption{Peculiar motions in the Perseus and Local arms.\footnotemark[$\ast$]}
\hline
\hline
\small
Source 	&l		&b		&$R$	&$\beta$\footnotemark[$\dag$] &$U$				&$V$				&$W$							&Arm	&Ref.\footnotemark[$\S$]	\\
		&deg	&deg	&kpc						&deg							&km s$^{-1}$	&km s$^{-1}$		&km s$^{-1}$					&	&	\\
\hline 
\endhead
\hline
\endfoot

\multicolumn{10}{l}{\hbox to 0pt{\parbox{180mm}{\footnotesize
\footnotemark[$\ast$] $R_{0}$ = 8.33 kpc (Gillessen et al. 2009), $\Theta_{0}$ = 240 km s$^{-1}$ (Reid $\&$ Brunthaler 2004), and a flat rotation curve [$\Theta$($R$) = $\Theta_{0}$] were assumed to determine the peculiar motions.\\
\footnotemark[$\dag$]Galactocentric azimuth [deg] (see text).\\
\footnotemark[$\ddag$] W49N and G48.60+0.02, located in the Perseus arm of the first Galactic quadrant, were excluded from the averaging (see text).\\
\footnotemark[$\S$]References: (1)This paper; (2)Ando et al. 2011; (3)Asaki et al. 2010; (4)Choi et al. 2008; Zhang et al. 2012a; (5)Choi et al. 2014; (6)Dzib et al. 2010; (7)Hirota et al. 2007; Kim et al. 2008; Menten et al. 2007; (8)Hirota et al. 2008a; (9)Hirota et al. 2008b; (10)Hirota et al. 2011; (11)Honma et al. 2012; (12)Imai et al. 2012; (13)Kusuno et al. 2013; (14)Loinard et al. 2008; (15)Moscadelli et al. 2009; (16)Moscadelli et al. 2011; (17)Nagayama et al. 2011; (18)Niinuma et al. 2011; (19)Oh et al. 2010; (20)Reid et al. 2009b; (21)Rygl et al. 2010; (22)Rygl et al. 2012;  (23)Sakai et al. 2012; (24)Sch\rm{\"o}nrich, Binney, $\&$ Dehnen 2010; (25)Shiozaki et al. 2011; (26)Torres et al. 2007; (27)Torres et al. 2009; (28)Xu et al. 2006; (29)Xu et al. 2009; (30)Xu et al. 2013; (31)Zhang et al. 2012b; (32)Zhang et al. 2013;
}}}
\endlastfoot

EC 95			&31.56	&5.33		&7.98		&1.6	&$-$1$\pm$4	&2$\pm$3	&2$\pm$0.5					&Local&6	 \\

W49N			&43.17	&0.01	&7.60$^{+0.55}_{-0.44}$		&88.3$^{+4.2}_{-4.2}$	&$-$17$\pm$14	&$-$24$\pm$13	&$-$5$\pm$14				&Perseus&32	\\
G48.60+0.02		&48.61	&0.02	&8.16$^{+0.41}_{-0.34}$		&81.4$^{+3.1}_{-3.1}$	&$-$5$\pm$12	&12$\pm$11		&6$\pm$12					&Perseus&32	\\

G59.78+0.06		&59.78	&0.06	&7.48$^{+0.02}_{-0.03}$		&14.4$^{+0.7}_{-0.7}$	&1$\pm$2	&$-$1$\pm$3	&$-$4$\pm$1				&Local&29	\\
ON1				&69.54	&$-$0.98	&7.82$^{+0.01}_{-0.01}$		&17.2$^{+1.5}_{-1.3}$	&7$\pm$5	&$-$6$\pm$5	&5$\pm$5			&Local	&17,21,30	\\

G74.03$-$1.71	&74.03	&$-$1.71	&8.04						&11.0$^{+0.3}_{-0.3}$	&8$\pm$2	&$-$6$\pm$5	&10$\pm$2				&Local&30	\\

G75.76+0.33		&75.76	&0.34		&8.20$^{+0.06}_{-0.04}$		&24.5$^{+2.0}_{-1.7}$	&0$\pm$6	&$-$16$\pm$9	&6$\pm$5				&Local&30	\\

ON2N			&75.78	&0.34	&8.24$^{+0.07}_{-0.05}$		&25.7$^{+2.2}_{-1.9}$	&$-$3$\pm$6	&$-$5$\pm$5	&1$\pm$5				&Local&2,30	\\
G76.38$-$0.61	&76.38	&$-$0.62	&8.12$^{+0.01}_{-0.01}$		&8.9$^{+0.7}_{-0.6}$	&3$\pm$18	&$-$11$\pm$5	&12$\pm$19				&Local&30	\\

IRAS 20126+4104	&78.12	&3.63		&8.15						&11.3$^{+0.4}_{-0.4}$	&6$\pm$1	&$-$13$\pm$5	&15$\pm$1				&Local&16	\\
AFGL 2591		&78.89	&0.71		&8.35$^{+0.02}_{-0.02}$		&23.0$^{+0.8}_{-0.7}$	&$-$13$\pm$4	&$-$11$\pm$7	&$-$22$\pm$5		&Local&22	\\
IRAS 20290+4052	&79.74	&0.99		&8.20						&9.4$^{+0.9}_{-0.7}$	&2$\pm$3	&$-$10$\pm$5	&6$\pm$3				&Local&22	\\
G79.87+1.17		&79.88	&1.18		&8.20						&11.2$^{+0.5}_{-0.5}$	&9$\pm$10	&$-$11$\pm$10	&3$\pm$10				&Local&30	\\
NML Cyg			&80.80	&$-$1.92	&8.23$^{+0.006}_{-0.003}$	&11.2$^{+0.9}_{-0.8}$	&$-$2$\pm$4	&$-$9$\pm$3		&$-$5$\pm$4				&Local&31	\\
DR 20			&80.86	&0.38		&8.23						&10.1$^{+0.6}_{-0.5}$	&7$\pm$2	&$-$9$\pm$5		&5$\pm$2				&Local&22	\\
DR 21			&81.75	&0.59		&8.25						&10.4$^{+0.6}_{-0.5}$	&0$\pm$2	&$-$8$\pm$3		&7$\pm$2				&Local&22	\\
W75N			&81.87	&0.78		&8.25						&8.9$^{+0.5}_{-0.5}$	&0$\pm$1	&2$\pm$3		&1$\pm$1				&Local&22	\\
G90.21+2.32		&90.21	&2.32		&8.36						&4.6$^{+0.1}_{-0.1}$	&$-$4$\pm$10	&$-$6$\pm$5		&6$\pm$10			&Local&30	\\
G92.67+3.07		&92.67	&3.07		&8.56$^{+0.01}_{-0.01}$						&11.0$^{+0.4}_{-0.3}$	&$-$16$\pm$3	&$-$6$\pm$10		&$-$2$\pm$2			&Local&30	\\
AFGL 2789		&94.60	&$-$1.80	&9.10$^{+0.14}_{-0.11}$		&19.6$^{+1.8}_{-1.5}$	&5$\pm$5	&$-$28$\pm$3	&$-$11$\pm$6				&Perseus&19	\\
G094.60-1.79	&94.60	&$-$1.80	&9.50$^{+0.21}_{-0.16}$		&24.5$^{+2.1}_{-1.8}$	&12$\pm$7	&$-$20$\pm$6	&$-$15$\pm$8				&Perseus&5	\\
G095.29-0.93	&95.30	&$-$0.94	&10.02$^{+0.10}_{-0.09}$		&28.8$^{+0.8}_{-0.8}$	&$-$9$\pm$5	&$-$7$\pm$5	&2$\pm$5				&Perseus&5	\\
G100.37-3.57	&100.38	&$-$3.58	&9.57$^{+0.11}_{-0.09}$		&20.8$^{+1.0}_{-0.9}$	&13$\pm$10	&$-$3$\pm$10	&3$\pm$10				&Perseus&5	\\

G105.41+9.87	&105.42	&9.88		&8.60$^{+0.02}_{-0.02}$		&5.6$^{+0.3}_{-0.3}$	&8$\pm$10	&$-$1$\pm$6		&$-$13$\pm$10			&Local&30	\\
IRAS 22198+6336	&107.30	&5.64		&8.59$^{+0.01}_{-0.01}$		&4.8$^{+0.2}_{-0.2}$	&5$\pm$2	&$-$7$\pm$5		&11$\pm$1			&Local&9,11	\\
L 1206			&108.18	&5.52	&8.60$^{+0.04}_{-0.03}$		&4.9$^{+0.6}_{-0.5}$	&0$\pm$3	&$-$8$\pm$3	&0$\pm$6				&Local&21	\\%
G108.20+0.58	&108.21	&0.59		&10.57$^{+0.59}_{-0.40}$		&23.3$^{+3.3}_{-2.6}$	&$-$11$\pm$10	&$-$15$\pm$9	&9$\pm$11				&Perseus&5	\\

IRAS 22480+6002		&108.43	&0.89		&9.42$^{+0.15}_{-0.12}$		&14.6$^{+0.8}_{-0.8}$	&12$\pm$3	&$-$23$\pm$2	&0$\pm$3				&Perseus&12	\\
G108.47-2.81	&108.47	&$-$2.82	&9.84$^{+0.06}_{-0.06}$		&18.1$^{+0.5}_{-0.5}$	&20$\pm$7	&$-$13$\pm$6	&$-$9$\pm$7				&Perseus&5	\\
G108.59+0.49	&108.59	&0.49		&9.41$^{+0.12}_{-0.10}$		&14.4$^{+1.1}_{-1.0}$	&46$\pm$6	&$-$5$\pm$6	&0$\pm$6				&Perseus&5	\\

Cep A			&109.87	&2.11	&8.59$^{+0.02}_{-0.02}$		&4.4$^{+0.3}_{-0.2}$	&2$\pm$4	&$-$2$\pm$5	&$-$5$\pm$2				&Local&15	\\
G111.23-1.23	&111.23	&$-$1.24&10.03$^{+0.82}_{-0.43}$		&18.0$^{+5.0}_{-3.3}$	&33$\pm$17	&3$\pm$15	&$-$2$\pm$15				&Perseus&5	\\
G111.25-0.77	&111.26	&$-$0.77&10.04$^{+0.17}_{-0.14}$		&18.1$^{+1.2}_{-1.0}$	&0$\pm$7	&$-$11$\pm$7	&$-$15$\pm$9				&Perseus&5	\\

NGC 7538		&111.54	&0.78	&9.62$^{+0.07}_{-0.07}$		&14.8$^{+0.6}_{-0.5}$	&19$\pm$2	&$-$23$\pm$3	&$-$10$\pm$2				&Perseus&15	\\
PZ Cas			&115.06	&$-$0.05&9.85$^{+0.14}_{-0.12}$		&15.0$^{+1.0}_{-0.9}$	&17$\pm$3	&7$\pm$3	&$-$6$\pm$4				&Perseus&13	\\
L 1287			&121.29	&0.66	&8.85$^{+0.02}_{-0.02}$		&5.1$^{+0.2}_{-0.2}$	&10$\pm$3	&$-$10$\pm$4	&$-$3$\pm$2				&Local&21	\\

W3(OH)			&133.95	&1.06	&9.79$^{+0.03}_{-0.03}$		&8.3$^{+0.1}_{-0.1}$	&19$\pm$3	&$-$11$\pm$2	&1$\pm$2				&Perseus&28	\\
S Per			&134.62	&$-$2.20	&10.18$^{+0.08}_{-0.08}$		&9.7$^{+0.3}_{-0.3}$	&3$\pm$3	&$-$14$\pm$3	&$-$6$\pm$2				&Perseus&3	\\

L 1448			&158.06	&$-$21.42	&8.53$^{+0.02}_{-0.01}$		&0.5				&$-$16$\pm$6	&$-$15$\pm$4		&$-$4$\pm$4			&Local&10	\\
NGC 1333		&158.35	&$-$20.56	&8.54$^{+0.02}_{-0.01}$		&0.5				&$-$10$\pm$3	&0$\pm$2			&4$\pm$2			&Local&8	\\
Hubble 4		&168.84	&$-$15.52	&8.46						&0.2				&$-$12$\pm$2	&1$\pm$0.4		&$-$9$\pm$1			&Local&26	\\
IRAS 05168+3634		&170.66	&$-$0.25	&10.19$^{+0.21}_{-0.17}$		&1.7$^{+0.2}_{-0.1}$	&8$\pm$2	&$-$13$\pm$6	&$-$7$\pm$8				&Perseus&23\\
HP-Tau/G2 		&175.73	&$-$16.24	&8.48		&0.08	&$-$18$\pm$2	&$-$1$\pm$0.2		&0$\pm$1				&Local&27	\\
T-Tau/Sb 		&176.23	&$-$20.89	&8.47		&0.1	&$-$18$\pm$0.9	&12$\pm$0.1			&$-$2$\pm$0.4				&Local&14	\\

G183.72-3.66	&183.72	&$-$3.66		&9.91$^{+0.03}_{-0.03}$		&$-$0.6$^{+0.01}_{-0.01}$	&0$\pm$5	&1$\pm$10	&3$\pm$10				&Perseus&5	\\

IRAS 06061+2151	 	&188.79	&1.03	&10.33$^{+0.13}_{-0.12}$		&$-$1.7$^{+0.1}_{-0.1}$		&12$\pm$1	&$-$22$\pm$2	&$-$11$\pm$1				&Perseus&18	\\
IRAS 06058+2138	 	&188.95	&0.89	&10.07$^{+0.11}_{-0.10}$		&$-$1.6$^{+0.1}_{-0.1}$		&6$\pm$1	&$-$14$\pm$3	&4$\pm$2				&Perseus&19	\\
S252 			&188.95	&0.89	&10.41$^{+0.03}_{-0.03}$		&$-$1.8$^{+0.02}_{-0.02}$	&$-$2$\pm$3	&$-$9$\pm$0.6	&$-$2$\pm$0.2				&Perseus&20	\\
G192.16$-$3.84 	&192.16	&$-$3.84	&9.81$^{+0.10}_{-0.08}$		&$-$1.9$^{+0.1}_{-0.1}$		&3$\pm$3	&$-$8$\pm$2	&3$\pm$1				&Perseus&25	\\
S255 			&192.60	&$-$0.05	&9.89$^{+0.07}_{-0.06}$		&$-$2.0$^{+0.07}_{-0.08}$	&3$\pm$2	&3$\pm$12	&3$\pm$7				&Perseus&21	\\

Orion  			&209.01	&$-$19.39	&8.67						&$-$1.2						&$-$1$\pm$4	&$-$1$\pm$3	&6$\pm$3				&Local&7	\\
G229.57+0.15	&229.57	&0.15		&11.83$^{+0.23}_{-0.20}$		&$-$17.2$^{+0.6}_{-0.7}$	&15$\pm$12	&$-$13$\pm$14	&$-$10$\pm$15				&Perseus&5	\\

G232.62+1.00 	&232.62	&1.00	&9.44$^{+0.07}_{-0.07}$		&$-$8.1$^{+0.4}_{-0.4}$			&2$\pm$3	&$-$3$\pm$3	&1$\pm$2				&Local&20	\\
G236.81+1.98	&236.82	&1.98		&10.33$^{+0.20}_{-0.17}$		&$-$14.4$^{+0.9}_{1.0}$	&$-$7$\pm$7	&$-$2$\pm$7	&$-$3$\pm$7				&Perseus&5	\\

VY CMa 			&239.35	&$-$5.07	&8.98$^{+0.07}_{-0.06}$		&$-$6.4$^{+0.5}_{-0.6}$			&0$\pm$2	&$-$9$\pm$3	&$-$3$\pm$2				&Local&4	\\
G240.31+0.07	&240.32	&0.07		&11.90$^{+0.40}_{-0.33}$		&$-$22.9$^{+1.3}_{-1.4}$	&$-$11$\pm$6	&4$\pm$7	&$-$14$\pm$7				&Perseus&	5\\

IRAS 07427-2400 &240.32	&0.07	&11.97$^{+0.75}_{-0.54}$		&$-$23.1$^{+2.1}_{-2.5}$	&$-$8$\pm$8	&0$\pm$9	&1$\pm$11				&Perseus&1	\\
DoAr21/Ophiuchus 			&353.02	&16.98	&8.21							&$-$0.1						&2$\pm$5	&$-$9$\pm$1	&6$\pm$2				&Local&14	\\
S1/Ophiuchus 			&353.10	&16.89	&8.22$^{+0.01}_{-0.01}$				&$-$0.1						&6$\pm$5	&$-$2$\pm$1	&$-$4$\pm$2				&Local&14	\\
\textbf{Sun}	&-	&-		&-	&-	&$U_{\odot}$=11.1$^{+1}_{-1}$		&$V_{\odot}$=12.24$^{+2}_{-2}$			&$W_{\odot}$=7.25$^{+0.5}_{-0.5}$	&	&24	\\
\hline
Average-1  	&	&	&	&						&$-$1$\pm$1	&$-$6$\pm$1	&1$\pm$1				&Local&	\\
Average-2\footnotemark[$\ddag$]  	&	&	&	&&8$\pm$3	&$-$9$\pm$2	&$-$4$\pm$1			&Perseus&	\\
\hline

\end{longtable}
}


\clearpage
\restoregeometry

\begin{figure*}[tb]
 \begin{center} 
     \includegraphics[width=160mm, height=160mm]{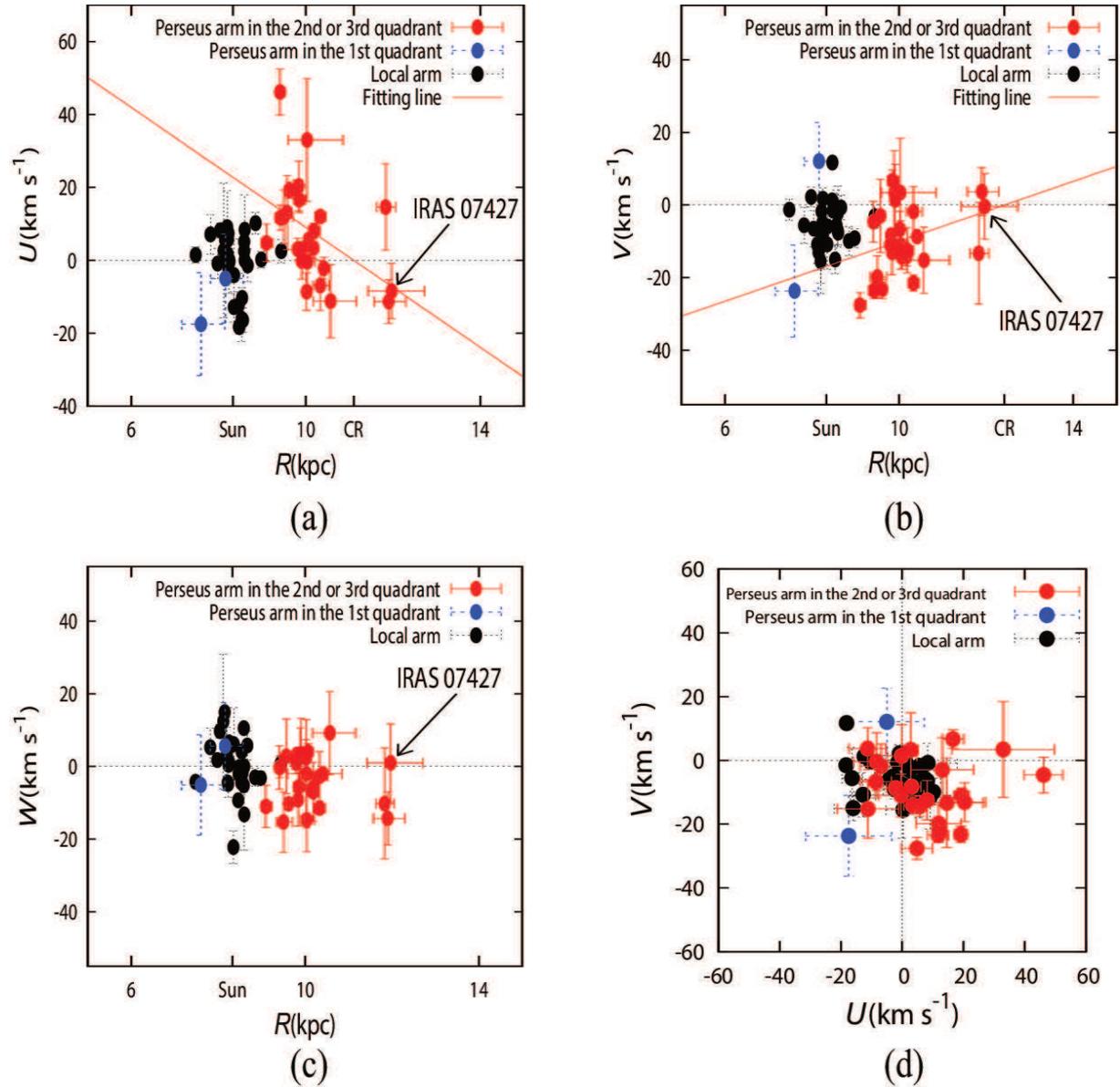}
\end{center}
\caption{(a)-(c) Peculiar motions ($U$, $V$, $W$) as a function of Galactocentric distance ($R$) shown in figures (a), (b) and (c), respectively. Red lines in (a) and (b) were fitted to the sources located in the Perseus arm of the second or third Galactic quadrant (see text). Red circles show the sources in the Perseus arm of the second or third Galactic quadrant, and blue ones display the sources in the Perseus arm of the first Galactic quadrant. Black circles represent the sources in the Local arm. (d) $U$ vs. $V$ plane.}
\label{fig5}
\end{figure*}
\clearpage

\begin{table*}[htb]
\begin{center}
\small
\caption{Summary of the asymmetric potential parameters used in this paper.}
\begin{tabular}{cll}
\hline
Parameter	&Dimension	&Notes\\ 
\hline
  $\sqrt[]{\Phi_{\rm{a}}}$	&(km s$^{-1}$)	&Amplitude of an asymmetric potential\\
  $R_{\rm{ref}}$	&(kpc)					&Reference position of an asymmetric potential\\
  $i$					&(deg)					&Pitch angle of an asymmetric potential\\
  $\Omega_{\rm{p}}$	&(km s$^{-1} \rm{kpc^{-1}}$)	&Pattern speed of an asymmetric potential\\
  $m$					&						&Mode of an asymmetric potential\\ 
$\lambda$			&(km s$^{-1} \rm{kpc^{-1}}$)	&Damping term for gas motion\\
$\varepsilon$		&(km s$^{-1} \rm{kpc^{-1}}$)	&Corotation softening parameter\\
    
\hline
\multicolumn{0}{@{}l@{}}{\hbox to 0pt{\parbox{180mm}{\footnotesize
\par\noindent
}\hss}}
\end{tabular}
\end{center}
\end{table*}

\begin{table*}[htb]
\begin{center}
\small
\caption{Results of the least-squares fits for the asymmetric potential parameters.\footnotemark[$\ast$]}
\begin{tabular}{ccccccccccl}
\hline
Model	&$\#$ of	&$\sqrt[]{\bf{\Phi_{\rm{a}}}}$\footnotemark[$\dag$]	&$R_{\rm{ref}}$				&{\bf i}	& $\Omega_{\rm{{\bf p}}}$										&{\bf m}	&$\lambda$					&$\varepsilon$	&$\chi^{2}$/d.o.f	&Memo\\
ID		&sources	&				&			&			&				&		&	&\\ 
\hline
1		&27			&{\bf 15.3}					&9.46$\pm$0.01			&{\bf 5.0}						&24.1$\pm$0.1							&{\bf 2}								&27.0$\pm$2.5			&0.99$\pm$0.03			&549/50		&30$\%$ of $\Sigma_{\odot}$\\
2		&27			&{\bf 28.1}					&9.39$\pm$0.02			&{\bf 10.0}						&17.8$\pm$0.8							&{\bf 2}								&24.4$\pm$1.9			&3.2$\pm$0.2			&339/50		&50$\%$ of $\Sigma_{\odot}$\\	
3		&27			&{\bf 34.7}					&9.10$\pm$0.04			&{\bf 15.0}						&17.9$\pm$0.4							&{\bf 2}								&18.9$\pm$1.7			&4.2$\pm$0.1			&321/50		&50$\%$ of $\Sigma_{\odot}$\\
4		  &27		  &{\bf 36.2}				&9.16$\pm$0.05		  	&{\bf 20.0}					  &10.1$\pm$0.2						      &{\bf 2}								  &6.1$\pm$0.3			  &13.2$\pm$0.3			  &319/50		  &40$\%$ of $\Sigma_{\odot}$\\
5		&27			&{\bf 12.5}					&9.74$\pm$0.01			&{\bf 5.0}						&24.1$\pm$0.1							&{\bf 4}								&7.7$\pm$0.6			&0.98$\pm$0.03			&684/50		&40$\%$ of $\Sigma_{\odot}$\\
6		&27			&{\bf 19.8}					&9.22$\pm$0.01			&{\bf 10.0}						&10.1$\pm$0.3							&{\bf 4}								&1.7$\pm$0.2			&8.8$\pm$0.3			&446/50		&50$\%$ of $\Sigma_{\odot}$\\	
7		&27			&{\bf 24.6}					&9.01$\pm$0.02			&{\bf 15.0}						&11.5$\pm$0.3							&{\bf 4}								&5.5$\pm$0.3			&9.3$\pm$0.3			&349/50		&50$\%$ of $\Sigma_{\odot}$\\
8		  &27		  &{\bf 25.6}				&9.48$\pm$0.03		  	&{\bf 20.0}					  &16.0$\pm$0.1						      &{\bf 4}								  &7.0$\pm$0.3			  &7.0$\pm$0.2			  &306/50		  &40$\%$ of $\Sigma_{\odot}$\\

\hline
\multicolumn{11}{@{}l@{}}{\hbox to 0pt{\parbox{150mm}{\footnotesize
\par\noindent \\
\footnotemark[$\ast$] Bold values, $\sqrt[]{\bf{\Phi_{\rm{a}}}}$, {\bf i}, and {\bf m}, are fixed based on previous researches (see text). Other parameters are searched within realistic values (see also text). $R_{\rm{ref}}$ is searched in a range corresponding to the phase between $-\pi$ and $\pi$ (rad) of the spiral potential from $R_{\rm{ref}}$ = 10.4 kpc with an increment of 0.1 kpc (see text). $\Omega_{\rm{{\bf p}}}$ is searched between 10 and 30 (km s$^{-1}$ kpc$^{-1}$) with an increment of 0.5 (km s$^{-1}$ kpc$^{-1}$). $\lambda$ and $\varepsilon$ are searched in the same range between 1 and 30 (km s$^{-1}$ kpc$^{-1}$) with an increment of 0.5 (km s$^{-1}$ kpc$^{-1}$). \\
\footnotemark[$\dag$]$\sqrt[]{\bf{\Phi_{\rm{a}}}}$ can be converted to surface density ($\Sigma$). The converted values are listed in the last column (see text).\\  
}\hss}}

\end{tabular}
\end{center}
\end{table*}

\subsection{Observational indication of the density-wave theory} 

Using the obtained peculiar motions and previous VLBI astrometry results, we show $U$ and $V$ of the Perseus and Local arms as a function of Galactocentric distance in figures 5a and 5b, respectively. Interestingly, linear trends might be seen in figures 5a and 5b for the Perseus arm, but except for W49N and G48.60+0.02 located in the Perseus arm of the first Galactic quadrant. For W49N and G48.60+0.02, we do not have a clear explanation for the deviations from the linear trends. As noted in the previous section, more astrometry results for the Perseus arm in 28.8 deg $\leq$ $\beta$ $\leq$ 81.4 deg, corresponding to 8.16 kpc $\leq$ $R$ $\leq$ 10.02 kpc (see table 5), are crucial for explaining the deviations.
 
According to Russeil (2007), the linear trend of $V$ components was previously confirmed in stellar rotation velocities (see fig. 7 in Russeil 2007), which is basically consistent with our result of the gaseous $V$ components. Russeil (2007) regarded Galactic corotation radius as the place where $V$ component is 0 km s$^{-1}$, since the density-wave theory predicted that directions of the peculiar motions ($U$ and $V$) are changed inversely at inner and outer corotation radii (e.g., Mel'nik et al. 1999). Based on the assumption, Russeil (2007) determined a corotation radius of 12.7 kpc by a linear equation fit toward the stellar $V$ components. In the same way, we conduct the linear fits toward both $U$ and $V$ components, but with W49N and G48.60+0.02 excluded from the fits. As a result, we obtain corotation radii of 11.1 $\pm$ 0.3 and 12.4 $\pm$ 0.5 kpc in $U$ and $V$ components, respectively. These VLBI astrometry results can be understood with the assumption of the density-wave theory, although more astrometry observations around the corotation radius are required to accurately determine the corotation radius.

To evaluate whether the astrometry results for the Perseus arm are affected by systematic errors (e.g., the assumed Galactic constants, peculiar solar motions, and Galactic rotation curve), we compare the peculiar motions of the Perseus arm with those of the Local arm as shown in table 5 and figure 5. Averaged disk peculiar motions of ($U_{{\rm mean}}$, $V_{{\rm mean}}$) = (8 $\pm$ 3, $-$9 $\pm$ 2) km s$^{-1}$ are obtained for 25 sources in the Perseus arm (W49N and G48.60+0.02 are excluded from the averaging), which are deviated from those of ($U_{{\rm mean}}$, $V_{{\rm mean}}$) = ($-$1 $\pm$ 1, $-$6 $\pm$ 1) km s$^{-1}$ for 32 sources in the Local arm with a significance of $\sim$ 1-$\sigma$. Note that the obtained errors are the standard errors. The difference between the Perseus and Local arms means that relative offsets between them in $U$ and $V$ components are real, although absolute values of $U$ and $V$ can be changed depending on the systematic errors. Figure 5d also shows the relative offsets between the Perseus and Local arms in $U \ {\rm vs.} \ V$ plane. Most sources in the Perseus arm of the second Galactic quadrant are located on the lower right side of the figure (see also table 5), while most sources in the Local arm are located around the origin, except for some outliers showing $U$ $\sim$ $-$20 km s$^{-1}$.    

To explain the peculiar motions of the Perseus arm in detail, we will compare these with an analytic gas dynamics model in the following section.

\begin{figure*}[tb]
 \begin{center} 
     \includegraphics[width=160mm, height=165mm]{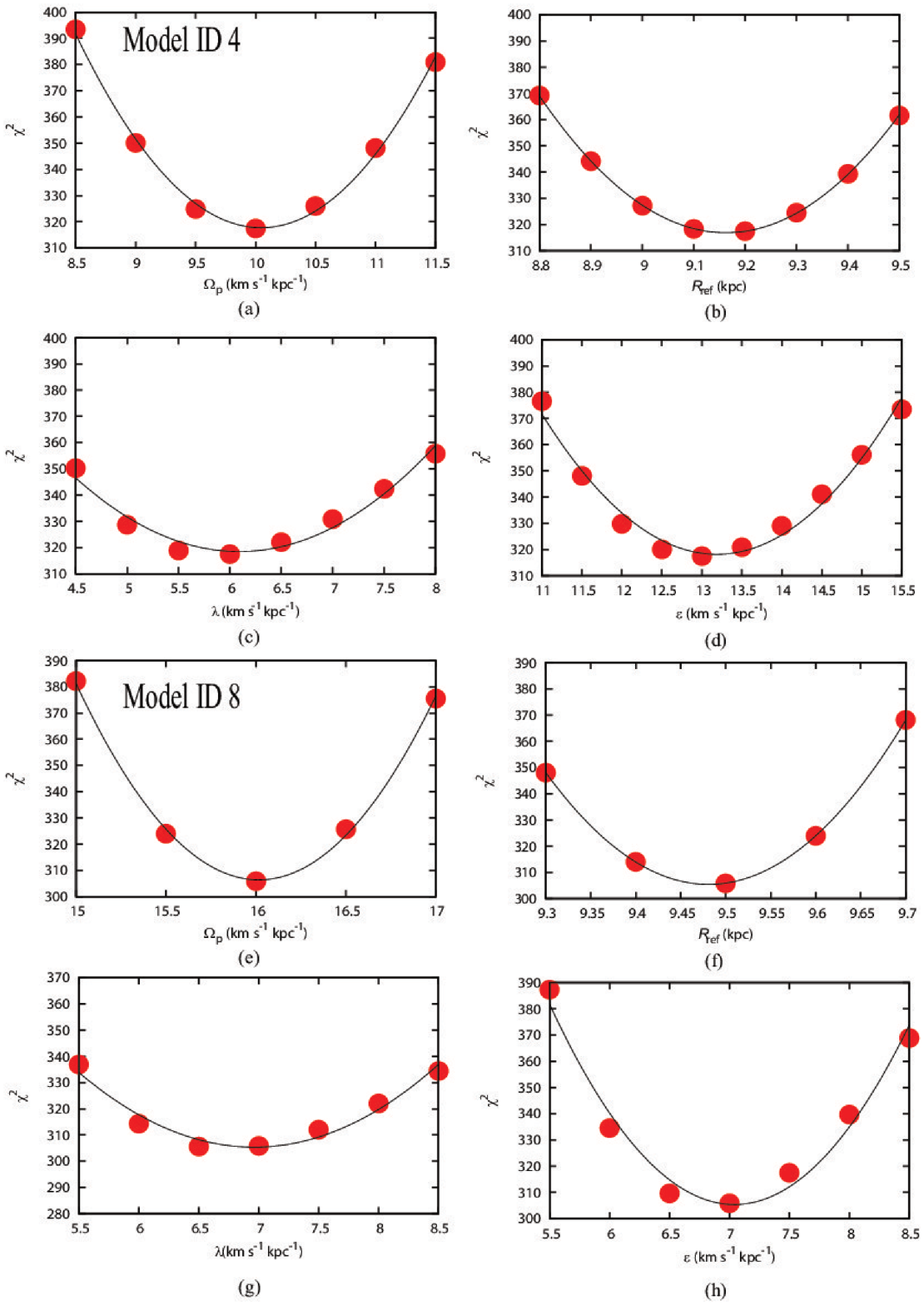}
\end{center}
\caption{\vtop{\hsize=400pt  $\chi^{2}$ distributions for each model parameter listed in table 7. (a)-(d) The $\chi^{2}$ distributions for model ID 4. (e)-(h) The $\chi^{2}$ distributions for model ID 8.  }}
\label{fig6}
\end{figure*}

\begin{figure*}[tb]
 \begin{center} 
     \includegraphics[width=152mm, height=155mm]{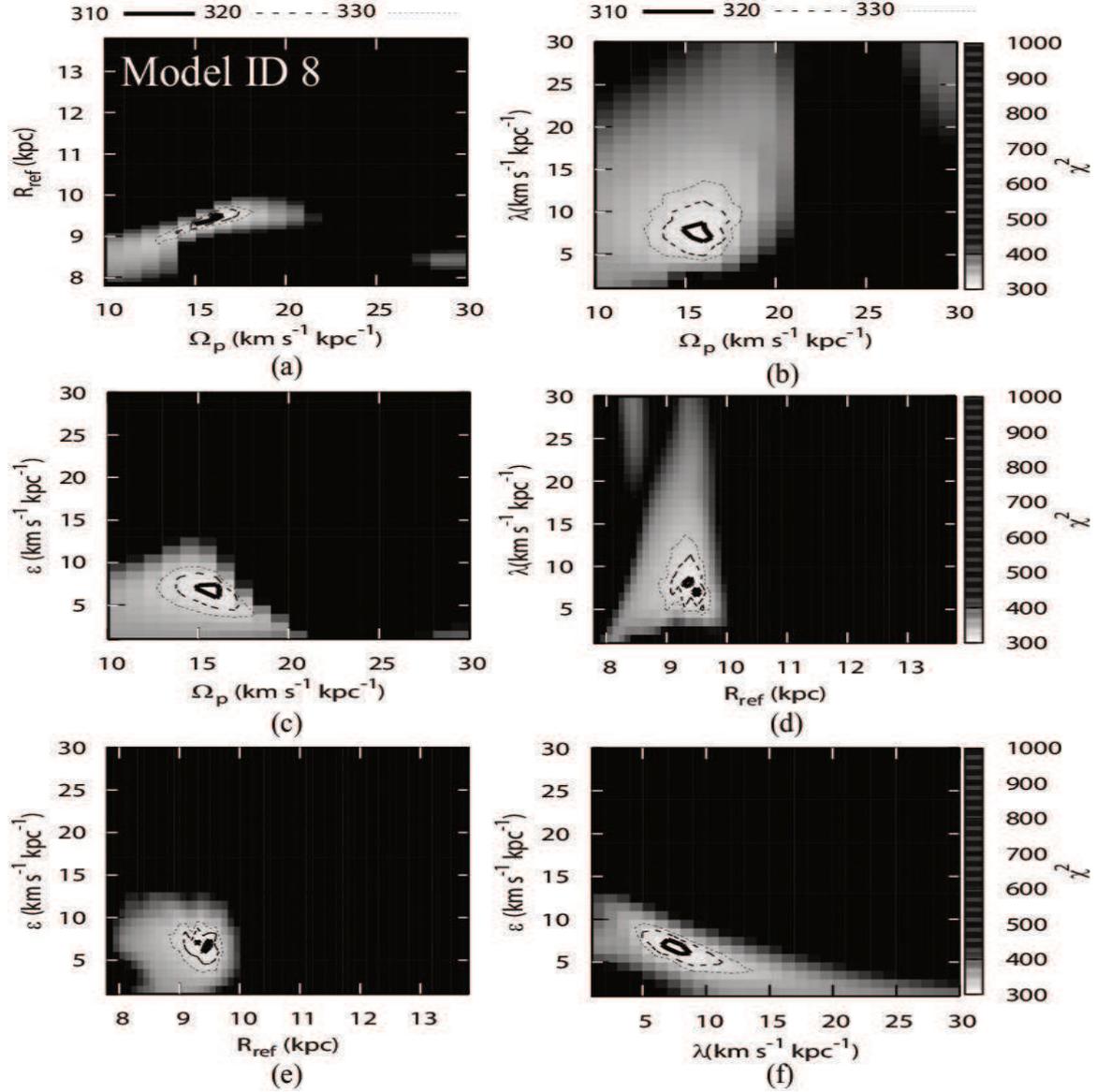}
\end{center}
\caption{\vtop{\hsize=400pt  $\chi^{2}$ distributions in various combinations of the model parameters listed in table 7, especially for model ID 8. Solid, dashed and dotted contours represent $\chi^{2}$ values of 310, 320 and 330, respectively. }}
\label{fig7}
\end{figure*}

\begin{figure*}[tb]
 \begin{center} 
     \includegraphics[width=163mm, height=160mm]{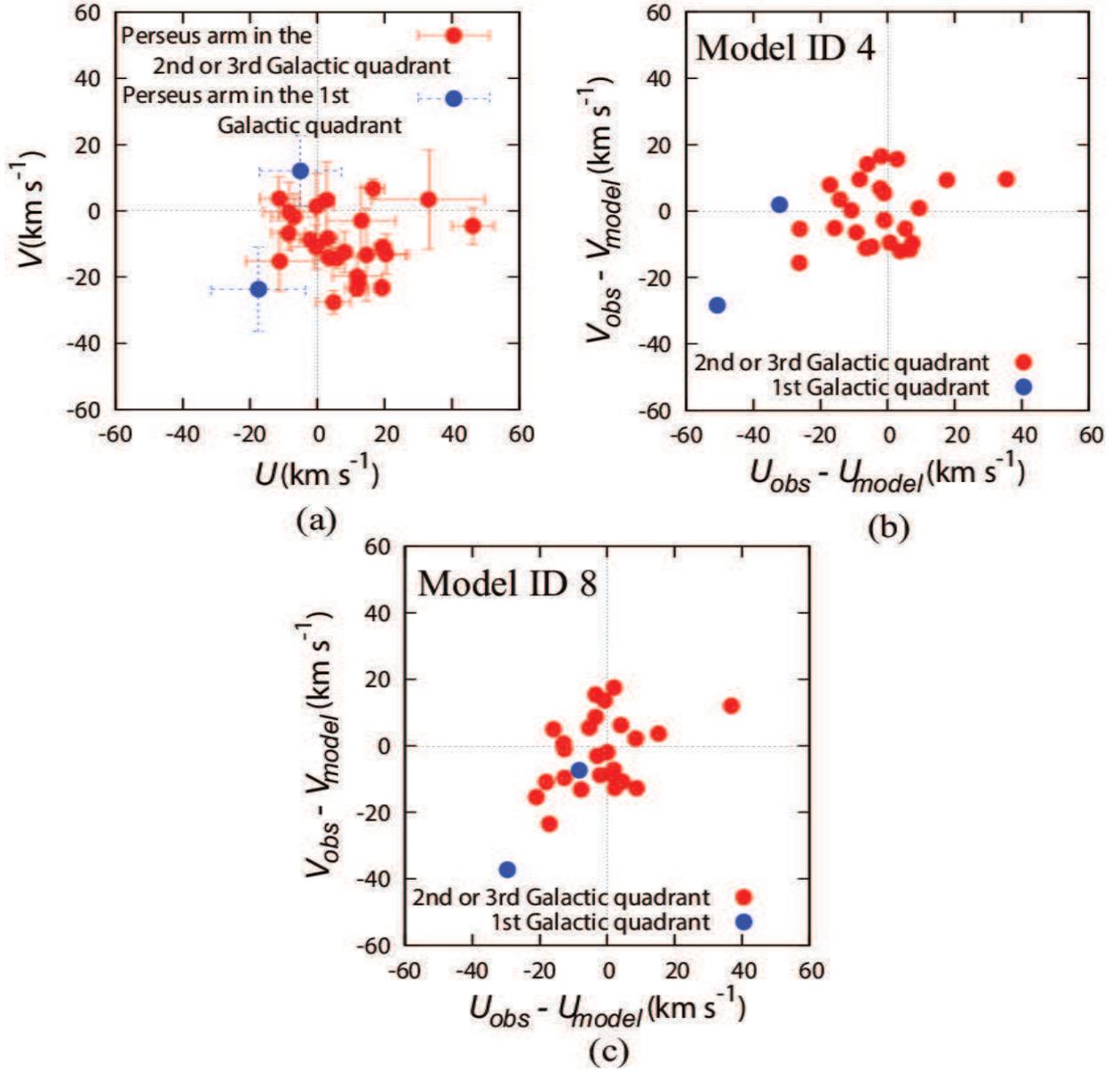}
\end{center}
\caption{(a) Peculiar motions plane ($U$ vs. $V$) for the sources in the Perseus arm (see table 5). Red circles show the sources in the Perseus arm of the second or third Galactic quadrant, and blue ones display the sources in the Perseus arm of the first Galactic quadrant. (b) Residuals between the observed peculiar motions and the models generated from ID 4 in table 7. (c) Same as (b), but from ID 8.}
\label{fig8}
\end{figure*}

\begin{figure*}[tb]
 \begin{center} 
     \includegraphics[width=149.0mm, height=140.0mm]{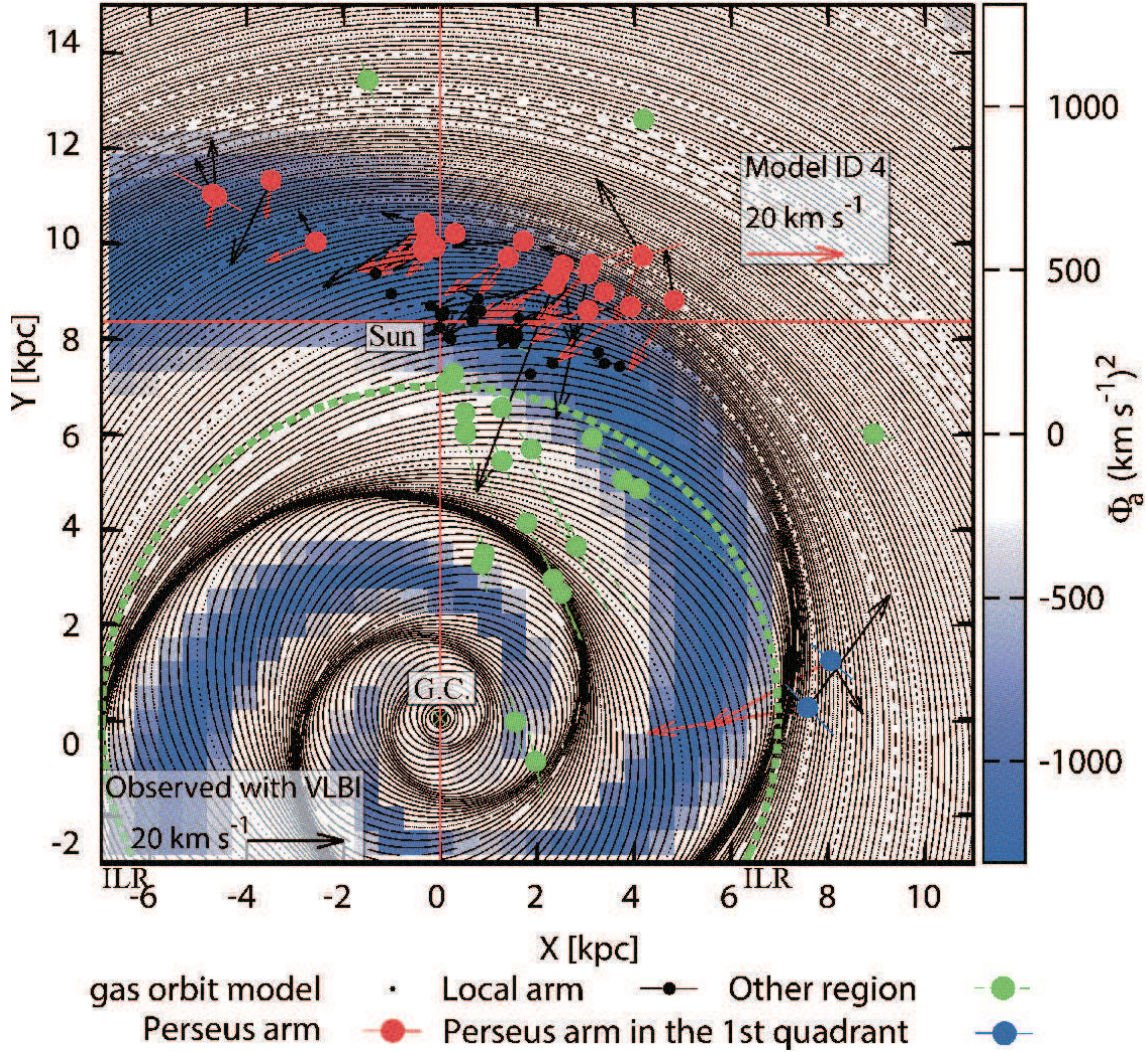}
\end{center}
\caption{The gas orbits model generated from model ID 4 listed in table 7. Filled circles show previous VLBI astrometry results, and different colors mean different regions (e.g., Red = Perseus arm in the second or third Galactic quadrant; Black = Local arm; Blue = Perseus arm in the first Galactic quadrant; Green = others). Black and red arrows show observed and modeled peculiar (non-circular) motions, respectively. Note that we assumed a flat rotation model (i.e., $\Theta(R)$ = $\Theta_{0}$), Galactic constants of ($R_{0}$, $\Theta_{0}$) = (8.33 kpc, 240.0 km s$^{-1}$), and peculiar solar motions of ($U_{\odot}$, $V_{\odot}$, $W_{\odot}$) = (11.1 $\pm$ 1, 12.24 $\pm$ 2, 7.25 $\pm$ 0.5) km s$^{-1}$ to derive the peculiar motions. Color bar represents amplitude of the spiral potential model (ID 4). Green dotted circle shows the Inner Lindblad Resonance (ILR). Solar position is (X, Y) = (0, $R_{0}$) kpc.}
\label{fig9}
\end{figure*}

\begin{figure*}[tb]
 \begin{center} 
     \includegraphics[width=151.0mm, height=143.0mm]{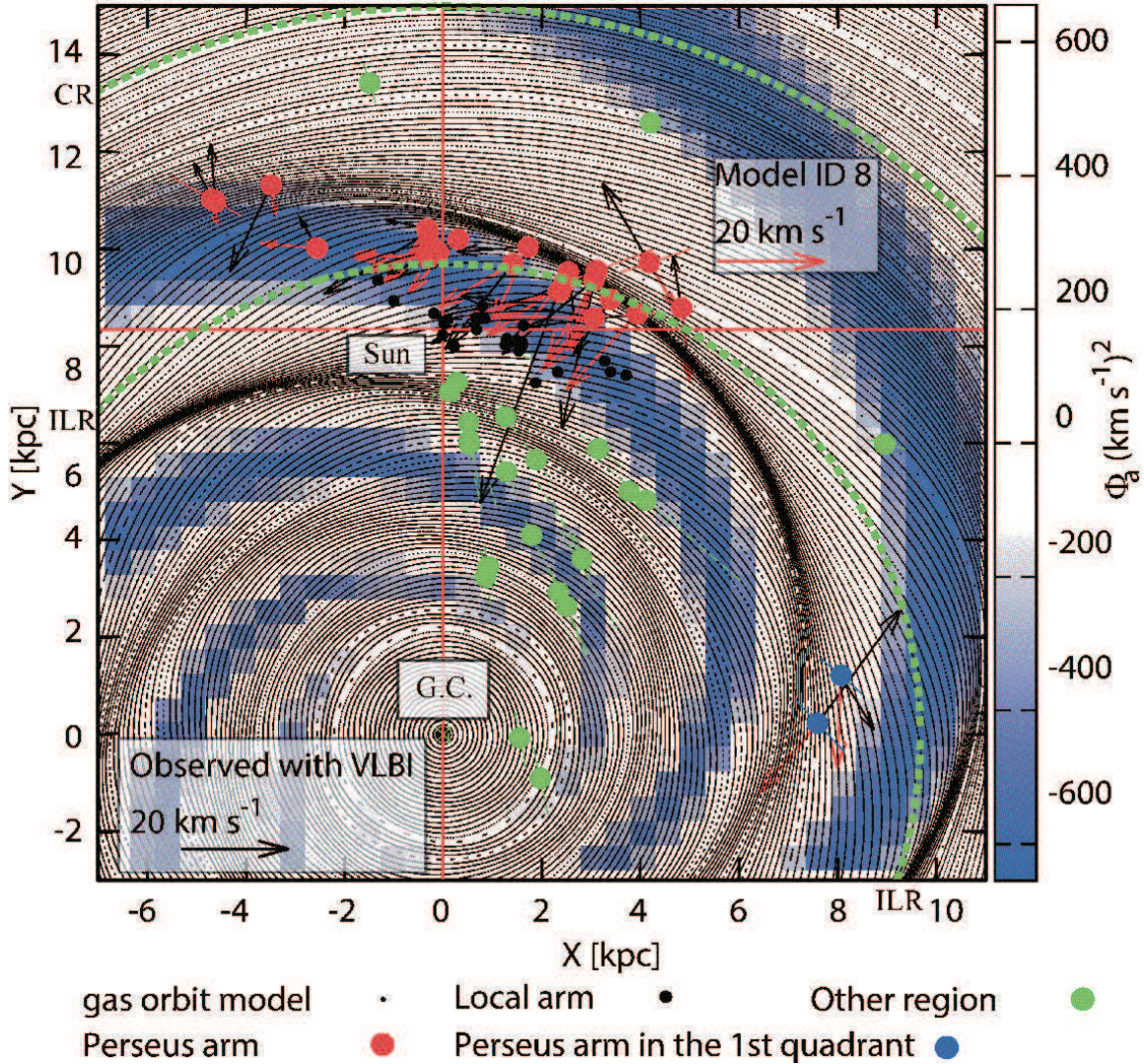}
\end{center}
\caption{Same as fig. 9, but using model ID 8 instead of ID 4. Inner and outer dotted circles represent the Inner Lindblad Resonance (ILR) and the Corotation Resonance (CR), respectively. }
\label{fig10}
\end{figure*}

\begin{figure*}[tb]
 \begin{center} 
     \includegraphics[width=166mm, height=132mm]{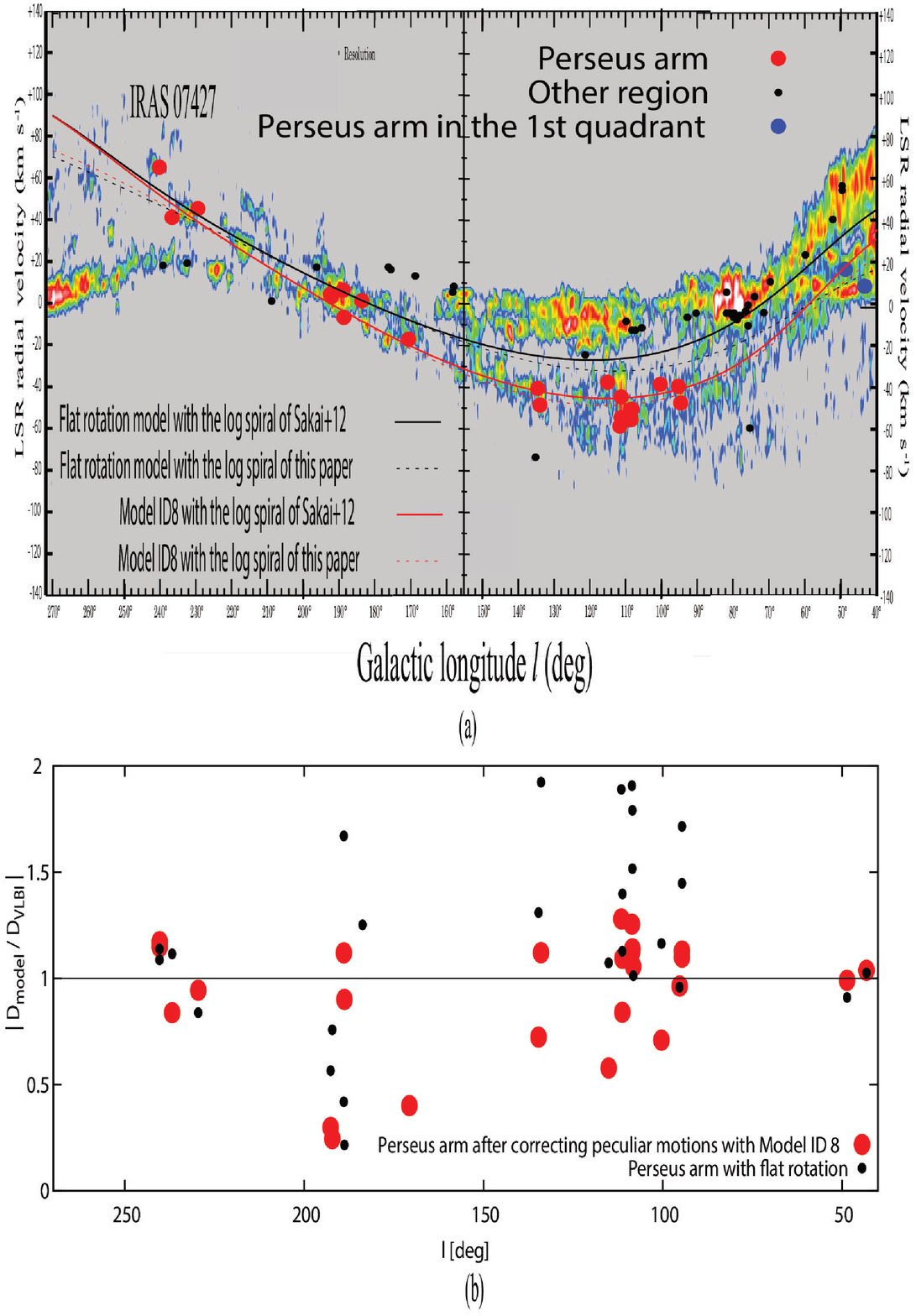}
\end{center}
\caption{ (a)VLBI astrometry results as colored circles superimposed on the longitude-velocity diagram of CO(J=1-0) emission (Dame et al. 2001). Different colors mean different regions (e.g., Red = Perseus arm in the second or third Galactic quadrant; Blue = Perseus arm in the first Galactic quadrant; Black = other regions). The horizontal axis shows Galactic longitude, and the vertical one represents the LSR velocity. The solid black curve is a model for tracing the Perseus arm, which is based on a flat rotation model (i.e., $\Theta(R)$ = $\Theta_{0}$) and the log spiral model (e.g., see eq. 1) of Sakai et al. (2012). The dashed black curve is the same as the solid one, but based on the log spiral model of this paper. Solid and dashed red curves are the same as the black curves, but with peculiar motions corrected based on model ID 8 in table 7 (see text). (b) Accuracies of distance measurements using the models (solid black and red curves) in fig. 11(a) relative to distance measurements by VLBI observations (see text). Horizontal axis shows Galactic longitude, and vertical one represents a ratio of both distance measurements, $|\frac{D_{\rm{model}}}{D_{{\rm VLBI}}}|$. Red and black circles were derived based on the solid red and black curves in fig. 11(a), respectively.} 
\label{fig11}
\end{figure*}

\subsection{Comparisons between observed peculiar motions and an analytic gas dynamics model}
Based on Pi$\tilde{\rm{n}}$ol-Ferrer et al. (2012) and (2014), we compare the observed peculiar (non-circular) motions listed in table 5 with an analytic gas dynamics model in this section. Detailed explanations for the model and the comparison procedure are summarized in {\bf Appendixes 2 and 3}. \\

\subsubsection{The model of the Milky Way}
Through the least-squares fits described in the {\bf Appendix 3}, we determine the reduced $\chi^{2}$ values (${\chi^{2}_{\nu}}$) in various combinations of model parameters as listed in tables 6 and 7. From models ID 1 to ID 4, the mode $m$ = 2 is fixed. The models ID 3 and ID 4 (in the cases of $i$ = 15.0$^{\circ}$ and 20.0$^{\circ}$) show lower $\chi^{2}_{\nu}$ values compared to ID 1 and ID 2. From ID 5 to ID 8, the mode $m$ = 4 is fixed, and ID 8 (in the case of $i$ = 20.0$^{\circ}$) shows the minimum $\chi^{2}_{\nu}$ value in the four models.     

Based on all the $\chi^{2}_{\nu}$ values, we regard the models ID 4 and ID 8 as good models in this paper. We note that these results are first step results, since we have to sophisticate both the fitting procedure (i.e., the least-squares fit) and the analytic model. For instance, we assumed the constant pitch angle and amplitude for the spiral potential, which may not represent the real Galaxy (e.g., a bifurcation or spur of the spiral arm). However, the sophistication of the model is beyond the scope of this paper.  

Figure 6 shows results of the least-squares fits especially for ID 4 and ID 8. Figure 7 checks convergences of the model parameters for ID 8. $\lambda$ (gas friction term) is not converged well in figure 7. However, $\lambda$ is just an artificial parameter and not an observable parameter. Figures 8b and 8c show comparisons between observed peculiar motions (fig. 8a) and modeled $U$ $\&$ $V$ with ID 4 and ID 8 applied, respectively. Averaged peculiar motions of ($U_{{\rm mean}}$, $V_{{\rm mean}}$) = (8 $\pm$ 3, $-$9 $\pm$ 2) km s$^{-1}$ are determined for 25 sources in the Perseus arm in fig. 8a (W49N and G48.60+0.02 are excluded from the averaging). Using the models ID 4 and ID 8, the systematic peculiar motions are canceled out as shown in figures 8b and 8c, respectively. In fact, averaged residuals as ($\overline{U_{{\rm obs}}-U_{{\rm model}}}$, $\overline{V_{{\rm obs}}-V_{{\rm model}}}$) are determined to be ($-$3 $\pm$ 3, 0 $\pm$ 2) km s$^{-1}$ and ($-$2 $\pm$ 2, $-$2 $\pm$ 2) km s$^{-1}$ with ID 4 and ID 8 applied, respectively. 
  
Figures 9 and 10 display the gas orbit models generated from ID 4 and ID 8, respectively. Observed and modeled peculiar motions for the Perseus arm are superimposed on the gas orbit models (see figures 9 and 10), and also amplitudes of the spiral potentials (ID 4 and ID 8) are represented in the figures. Interestingly, the gas orbits of ID 8 concentrate around the sources located in the Perseus arm, meaning that our model (ID 8) based on the observed peculiar motions can make a dense gas region precisely.  \\

\subsubsection{Offset between gas and spiral potential model}
As clearly seen in fig. 10, a dense gas region traced by VLBI astrometry results is located downstream of the spiral potential model, which is consistent with Pi$\tilde{\rm{n}}$ol-Ferrer et al. (2012) and (2014), but not consistent with the galactic shock proposed by Fujimoto (1968) and  Roberts (1969). The difference of the gas distributions may originate in a difference between analytic (Pi$\tilde{\rm{n}}$ol-Ferrer et al. 2012 and 2014) and numerical (Fujimoto 1968; Roberts 1969) solutions.

Observationally, the dense gas region at the downstream was confirmed in Egusa et al. (2011). Egusa et al. (2011) conducted CO(J=1-0) observations with CARMA toward the nearby grand-design spiral galaxy M51. They found that massive Giant Molecular Clouds (GMCs) were located downstream of the spiral arm, while smaller clumps were located upstream of the spiral arm. In addition, the massive GMCc were traced by the brightest HII regions. 

Using a dynamical model of the Milky Way (fig. 10), we calculate the time needed to pass the offset between the dense gas region and the bottom of the spiral potential model for estimating the typical time scale of the star formation. The calculated time scale is 46 Myr, which is longer than a star-formation time-scale of $<$ 10 Myr proposed by Egusa et al. (2009). Also, 46 Myr is longer than the free fall time ($t_{ff}$) for GMCs and a lifetime of massive GMCs, which are $\sim$ 10 Myr (see fig. 1d in Koyama $\&$ Inutsuka 2000 with a number density of 10$^{2}$ cm$^{-3}$) and $\sim$ 30 Myr (Stahler $\&$ Palla 2005), respectively. Note that gas depletion times are much longer than $t_{ff}$ in the Galactic molecular clouds (e.g., see table 4 in Evans et al. 2009).

On the other hand, 46 Myr might be explained, if one assumes a magnetically supported cloud where a timescale of magnetic flux loss ($t_{B}$) is 10$-$100 times longer than the free fall time ($t_{B}$ $\gg$ $t_{ff}$) (see fig. 3 in Nakano et al. 2002). Observationally, Sato et al. (2008) determined a dynamical age of $\sim$ 20 Myr for a star-forming region associated with the NGC 281 superbubble with VLBI astrometry. 

Based on the above discussions, the longer time scale (46 Myr) derived from a dynamical model of the Milky Way might be explained, if star formation is triggered downstream of the spiral potential model, or if molecular clouds are magnetically supported.

\subsubsection{The ``revised kinematic distance'' with our model}
Here we model loci of the Perseus arm on the $l-v$ diagram of CO(J=1-0) emission (Dame et al. 2001) using the model ID 8 in table 7. Generally, the observed line-of-sight velocity ($V_{{\rm LSR}}$) can be written by  

\begin{equation}
V_{{\rm LSR}} = \left( \frac{\Theta}{R}-\frac{\Theta_{0}}{R_{0}}\right) R_{0}{\rm sin}l \\
\end{equation}
with the assumption of circular motion. Here $\Theta$ is the rotation speed at a Galactocentric distance of $R$, $l$ is Galactic longitude, and ($R_{0}$, $\Theta_{0}$) are the Galactic constants. We modify eq. 2 with correction terms about the peculiar motions as   
\begin{equation}
V_{{\rm LSR}} = \left( \frac{\Theta+V}{R}-\frac{\Theta_{0}}{R_{0}}\right) R_{0}{\rm sin}l \ - \ U{\rm cos}\phi,\\
\end{equation}
where $U$ and $V$ are the peculiar motions, and $\phi$ = 180.0$^{\circ}$ $-$ ($l$ + $\beta$). Note that $\beta$ is the Galactocentric azimuth explained previously. For $R$ we can use the logarithmic spiral models for the Perseus arm as determined in this paper and Sakai et al. (2012). For the peculiar motions ($U$ and $V$) we can refer to the model ID 8.  

As a result, fig. 11a shows the loci for the Perseus arm with or without the corrections of the peculiar motions. Note that we assumed a flat rotation model (i.e., $\Theta(R)$ = $\Theta_{0}$) in both procedures. The astrometry results listed in table 5 are also superimposed on fig. 11a. Clearly, the corrections with ID 8 work well to trace the sources in the Perseus arm on the $l-v$ diagram compared to the case without the corrections. The obtained loci (as red curves in fig. 11a) can be used to estimate a distance for a source located on the loci (i.e., the source located in the Perseus arm). 

To evaluate an accuracy of the distance measurement about for the loci (as red curves in fig. 11a), we formulate the revised kinematic distance ($D$) using eq. 3 as 
\begin{eqnarray*}
R = (\Theta \ + \ V)\left( \frac{R_{0}{\rm sin}l}{V_{{\rm LSR}}+U{\rm cos}\phi+\Theta_{0}{\rm sin}l}\right)\\
\end{eqnarray*}
and
\begin{equation}
D = R_{0}{\rm cos}l \pm \sqrt{R^{2}-R^{2}_{0}{\rm sin}^{2}l}. \\
\end{equation} 
Note that in the final step we just used the law of cosines. Using eq. 4 with the corrections of the peculiar motions, we can determine distances toward the sources located on the loci in fig. 11a. Ratios of the revised kinematic distances to VLBI astrometry results are shown as red circles in fig. 11b. 

On the other hand, the same procedures are conducted toward the ``general'' kinematic distances in fig. 11b (as shown by black circles). Especially for Galactic longitude $l$ between 94.60$^{\circ}$ (corresponding to AFGL 2789) and 134.62$^{\circ}$ (corresponding to S Per), our model (ID 8) shows good corrections, although around $l$ $\sim$ 180$^{\circ}$ it still has large uncertainties. The averaged ratio in the case of the corrected distances is 1.03 $\pm$ 0.20 with 94.60$^{\circ}$ $\leq$ $l$ $\leq$ 134.62$^{\circ}$ for 14 sources in the Perseus arm, while that in the case of the kinematic distances is 1.45 $\pm$ 0.35. Note that the errors are the standard deviations. \\ \\

\section{Conclusion}
We showed astrometry results of IRAS 07427-2400. Trigonometric parallax was obtained to be 0.185 $\pm$ 0.027 mas, corresponding to a distance of 5.41$^{+0.92}_{-0.69}$ kpc, which is consistent with the previous result of 0.188 $\pm$ 0.016 mas obtained by Choi et al. (2014) within error. Systematic proper motions were determined to be ($\mu_{\alpha}$cos$\delta$, $\mu_{\delta}$) = ($-$1.79 $\pm$ 0.32, 2.60 $\pm$ 0.17) mas yr$^{-1}$ with six maser features ranging 62.3 $\leq$ $V_{{\rm LSR}}$ $\leq$ 70.4 km s$^{-1}$, while Choi et al. (2014) showed ($\mu_{\alpha}$cos$\delta$, $\mu_{\delta}$) = ($-$2.43 $\pm$ 0.02, 2.49 $\pm$ 0.09) mas yr$^{-1}$ with one maser feature ranging 67.1 $\leq$ $V_{{\rm LSR}}$ $\leq$ 68.0 km s$^{-1}$. Averaging of the multiple features is essential to accurately estimate the systematic proper motions, since a difference of observed proper motions was up to $\sim$ 2 mas, corresponding to $\sim$ 50 km/s at a distance of 5.4 kpc.

Also, we succeeded in conducting direct (quantitative) comparisons between VLBI astrometry results and an analytic gas dynamics model for testing the density-wave theory, while the direct comparisons have suffered from observational limitations such as insufficient spatial and velocity resolutions since 1964, in which the density-wave theory was proposed by Lin $\&$ Shu (1964). We obtained mainly two results from the direct comparisons, although sophistication of the model should be conducted (e.g., a bifurcation of the spiral arm). 

First is the different pitch angles of gas and (probably) stellar spiral arms. The pitch angle of the Perseus arm was obtained to be 11.1 $\pm$ 1.4 deg by VLBI astrometry results in this paper, while that of the spiral potential model was obtained to be $\sim$ 20 deg. The difference of the pitch angles will be distinguishable by stellar astrometry (e.g., Gaia astrometry). In addition, a dense gas region traced by VLBI astrometry results was located downstream of the spiral potential model (see fig. 10), which was previously confirmed in the nearby grand-design spiral galaxy M51 in Egusa et al. (2011). The offset between the dense gas region and the spiral potential model will also be distinguished by stellar astrometry.  

In the Gaia era, a combination of gas and stellar astrometry will be a powerful tool to distinguish several dynamical models such as the density-wave theory and the recurrent transient spiral proposed by mainly numerical simulations. The model selection will allow us to understand the origin and evolution of the spiral arm, as well as those of the Milky Way. 

\bigskip

 \hspace*{30pt} We are grateful to VERA project members for the support they offered during observations.
We would like to thank the referee for carefully reading the manuscript. We would also like to thank Ms. Yolande McLean for conducting English proofreading. This work was financially supported by the National Astronomical Observatory of Japan (NAOJ) and the Grant-in-Aid for the Japan Society for the Promotion of Science Fellows (NS).


\appendix
\onecolumn
\section{Peculiar motions and those errors calculation}
Here, section 3.3 is explained in detail.

For Steps 2 and 3 (coordinates conversion and proper motion to absolute motion), we refer to the simple matrix multiplication shown by Johnson $\&$ Soderblom (1987) to determine full-space motions of the source at the solar position as below:
\[
\left[
    \begin{array}{c}
      U' \\
      V' \\
      W' 
    \end{array}
  \right]
= \bf{B} \cdot \left[
    \begin{array}{c}
      v_{\rm{helio}}\\
      k\mu_{\alpha}\rm{cos}\delta/\pi \\
      k\mu_{\delta}/\pi
    \end{array}
  \right]
\]
where
\[
  \bf{B} = \bf{T} \cdot \bf{A}
\]
with
\[
      \bf{T} = 
\left[
    \begin{array}{lll}
      +\rm{cos}\theta_{0}	&+\rm{sin}\theta_{0}&0\\
      +\rm{sin}\theta_{0}	&-\rm{cos}\theta_{0}&0\\
      0						&0					&+1
    \end{array}
    \right]
\left[
    \begin{array}{lll}
          -\rm{sin}\delta_{\rm{NGP}}	&0	&+\rm{cos}\delta_{\rm{NGP}}\\
          0								&-1	&0\\
          +\rm{cos}\delta_{\rm{NGP}}	&0	&+\rm{sin}\delta_{\rm{NGP}}\\
    \end{array}
  \right]
\left[
    \begin{array}{lll}
          +\rm{cos}\alpha_{\rm{NGP}}	&+\rm{sin}\alpha_{\rm{NGP}}	&0\\
          +\rm{sin}\alpha_{\rm{NGP}}	&-\rm{cos}\alpha_{\rm{NGP}}	&0\\
          0								&0							&+1\\
    \end{array}
  \right]
\]
and
\[
      \bf{A} = 
\left[
    \begin{array}{lll}
      +\rm{cos}\alpha \ \rm{cos}\delta	&-\rm{sin}\alpha 	&-\rm{cos}\alpha \ \rm{sin}\delta \\
      +\rm{sin}\alpha \ \rm{cos}\delta	&+\rm{cos}\alpha	&-\rm{sin}\alpha \ \rm{sin}\delta\\
      +\rm{sin}\delta					&0					&+\rm{cos}\delta
    \end{array}
    \right].
\]
Here k = 4.74057 and ($U'$, $V'$, $W'$) are 3D-velocity vectors at the solar position in Galactic coordinates. $U'$, $V'$, and $W'$ are directed toward the Galactic center, the Galactic rotation, and the North Galactic pole, respectively. We cite Reid et al. (2009a) to set right ascension and declination of the North Galactic pole ($\alpha_{\rm{NGP}}$, $\delta_{\rm{NGP}}$) in J2000 coordinates and the position angle of the North Celestial pole ($\theta_{0}$).   

 Using the above corrections, we can write the peculiar motions at the source position with matrix formulas below:

\[
\left[
    \begin{array}{c}
      U \\
      V \\
      W 
    \end{array}  
\right]
=
\bf{C} \cdot 
\left[
    \begin{array}{c}
      v_{\rm{helio}}\\
      k\mu_{\alpha}\rm{cos}\delta/\pi \\
      k\mu_{\delta}/\pi
    \end{array}
  \right]
  + \bf{D} \cdot
  \left[
    \begin{array}{c}
      U_{\odot}\\
      V_{\odot}\\
      W_{\odot}
    \end{array}
  \right]
  -
  \left[
    \begin{array}{c}
      0\\
      \Theta(R)\\
      0
    \end{array}
  \right]
  \]
where  
  \[
\bf{C} = \bf{D} \cdot \bf{B}
\]
and
\[
\bf{D} =
\left[
    \begin{array}{lll}
          +\rm{cos}\beta	&-\rm{sin}\beta	&0\\
          +\rm{sin}\beta	&+\rm{cos}\beta	&0\\
          0								&0							&+1\\
    \end{array}
  \right].
\]
Here $\beta$ is Galactocentric azimuth ($\beta$ is defined as 0 deg toward the Sun from the Galactic Center and increasing clockwise).

 For the error calculation of the peculiar motions, we show the matrix formula as
\[
\left[
    \begin{array}{c}
      \delta^{2}_{U} \\
      \delta^{2}_{V} \\
      \delta^{2}_{W} 
    \end{array}  
\right]
=
\bf{E}
\left[
    \begin{array}{c}
      \delta_{v_{\rm{helio}}}^{2}\\
       (k/\pi)^{2}[\delta_{\mu_{\alpha}\rm{cos}\delta}^{2} + (\mu_{\alpha}\rm{cos}\delta \ \delta_{\pi}/\pi)^{2}] \\
       (k/\pi)^{2}[\ \ \ \ \delta_{\ \mu_{\delta}}^{2} \ + \ \ \ \ \ (\mu_{\delta} \delta_{\pi}/\pi)^{2}\ \ \ \ ]
    \end{array}
  \right]
  + \rm{2}\mu_{\alpha}\rm{cos}\delta\mu_{\delta}{\it k}^{2}\sigma_{\pi}^{2}/\pi^{4}
  \left[
    \begin{array}{c}
      c_{12} \cdot c_{13} \\
      c_{22} \cdot c_{23} \\
      c_{32} \cdot c_{33}
    \end{array}
  \right].
\]
Here the elements of the matrix $\bf{E}$ are the squares of the individual elements of $\bf{C}$, i.e., $e_{ij}$ = $c_{ij}^{2}$ for all $i$ and $j$. Note that the final matrix can be applied at any place of the Galactic disk, while the matrix (eq. 2) in Johnson $\&$ Soderblom (1987) was derived with $\beta$ = 0 deg. Clearly, both matrices are identical if $\beta$ is 0 deg.

\onecolumn
\section{Introduction of the analytic gas dynamics model (Pi$\tilde{\rm{n}}$ol-Ferrer et al. 2012 and 2014)}
The model proposed by Pi$\tilde{\rm{n}}$ol-Ferrer et al. (2012) and (2014) solves the equations of motion using the linear epicyclic approximation (see Binney $\&$ Tremaine 2008, p.189), which is regarded as an extension of previous research related to galactic dynamics (e.g., Lindblad 1927; Lindblad 1958; Sanders $\&$ Huntley 1976; Lindblad $\&$ Lindblad 1994; Wada 1994). 

To explain the observed velocity ellipsoid in the Milky Way, Lindblad (1927) introduced the epicyclic description of nearly circular orbits for stars in a circularly symmetric galaxy. Lindblad (1958) added a perturbing potential, showing rigid-body rotation with the pattern speed ($\Omega_{p}$), in the theory given by Lindblad (1927). He pointed out that there were resonances in the perturbed system. The resonances are now called as the Lindblad resonances. Lindblad $\&$ Lindblad (1994), as well as Wada (1994), introduced gas dynamical friction ($\lambda$) in the epicyclic approximation to examine gas motions. Using the theory, they succeeded in examining the gas motions perturbed by a bar potential without the occurrence of the Lindblad resonances, thanks to the introduction of $\lambda$. However, another resonance called the corotation resonance (CR) occurred at the place where $\Omega_{p}$ = $\Omega(R)$ (as the angular speed of star and gas) in their papers, which forced them to research the gas motions in inner corotation radius (where $\Omega(R)$ $>$ $\Omega_{p}$) for avoiding the resonance: they examined the gas motions perturbed by a bar potential around galactic center where $\Omega(R)$ $>$ $\Omega_{p}$. 

Following the researches described above, Pi$\tilde{\rm{n}}$ol-Ferrer et al. (2012) and (2014) introduced artificial damping term ($\varepsilon$) with the gas dynamical friction $\lambda$ in the epicyclic approximation to examine gas motion in the entire disk without the all resonances (i.e., co-rotation and Lindblad resonances). The important point in the new model is that they formulated the model to be able to choose any asymmetric potentials such as bar and spiral potentials, while comparisons between observations and the previous model provided by Lindblad $\&$ Lindblad (1994) and Wada (1994) have been conducted in always bar potentials (e.g., Lindblad et al. 1996; Sakamoto et al. 1999; Boone et al. 2007). 

Using the new model provided by Pi$\tilde{\rm{n}}$ol-Ferrer et al. (2012) and (2014), one can obtain steady solutions for gas motions perturbed by an arbitrary asymmetric potential showing rigid-body rotation with $\Omega_{p}$ in a co-rotating flame. Mathematical descriptions for the model are summarized in the Appendix of Pi$\tilde{\rm{n}}$ol-Ferrer et al. (2012), and therefore here we just explain important parts of the model. First, we write axisymmetric and asymmetric potentials in polar coordinates ($R$, $\theta$) by 
\begin{equation}
\Phi(R, \theta) = \Phi_{0}(R) + \Phi_{1}(R, \theta), \\
\end{equation}
where subscripts 0 and 1 denote axisymmetric and asymmetric potentials, respectively. For a technical reason, we regard the Galactic center as the origin of the polar coordinates, and also $\theta$ is defined to be 90$^{\circ}$ toward the Sun and increasing counterclockwise in this paper. In the situation, $\theta$ is related to the Galactocentric azimuth $\beta$, which is $\theta$ = 90$^{\circ}$ $-$ $\beta$. 

Second, the asymmetric potential can be rewritten to be 
\begin{equation}
\Phi_{1}(R, \theta) = - \sum_{m=1}^{n} \Psi_{m}(R){\rm cos} \hspace{0.1em}m(\theta-\vartheta_{m}(R)). \nonumber \\
\end{equation}
Here $m$ is a mode of the asymmetric potential (e.g., $m$ = 2 for a bar potential), $\Psi_{m}$ is an amplitude of the asymmetric potential, and $\vartheta_{m}(R)$ shows a phase of the asymmetric potential: In a bar potential $\vartheta_{m}(R)$ is constant, and for a spiral potential here we assume the logarithmic-spiral model as explained in section 4.1. For $\Psi_{m}$($R$) here we assume it as constant for simplicity in the same manner as Sumi et al. (2009). In the case of the logarithmic-spiral model, $\vartheta_{m}(R)$ can be written by
\begin{equation}
\vartheta_{m}(R) = {\rm cot}\hspace{0.1em}i\hspace{0.1em}{\rm ln}\frac{R}{R_{{\rm ref}}} \ + \ 90.0^{\circ}.  \\
\end{equation}
Note that $i$ and $R_{{\rm ref}}$ are the pitch angle and the reference position in the logarithmic spiral model. 

Third, $R$ and $\theta$ in the polar coordinates are also decomposed into axisymmetric and asymmetric parts as 
\begin{equation}
R = r_{0} + \xi \\
\end{equation}
and
\begin{equation}
\theta = \theta_{0}+(\Omega-\Omega_{p})t+\frac{\eta}{r_{0}}. \\
\end{equation}
Here $\xi$ and $\eta$ are deviations from circular motion, $t$ is time, and $\Omega(r_{0})$ is an angular velocity of the circular motion. Note that the axisymmetric part $r_{0}$ is different from the Galactic constant $R_{0}$. Fourth, based on Pi$\tilde{\rm{n}}$ol-Ferrer et al. (2012) and (2014), the equations of motion (see also Binney $\&$ Tremaine 2008, p.189) are linearized by neglecting higher order terms of $\xi$ and $\eta$ as 
\begin{equation}
\ddot{\xi} + 2\lambda\dot{\xi}-2\Omega\dot{\eta}-4\Omega A \xi = -\frac{\partial \Phi_{1}}{\partial R}   \\
\end{equation}
and
\begin{equation}
\ddot{\eta} +2\Omega\dot{\xi} + 2\lambda\dot{\eta}+4\lambda A \xi = -\frac{1}{R}\frac{\partial \Phi_{1}}{\partial \theta}.   \\
\end{equation}
Provided that we are not close to the corotation (where $\Omega = \Omega_{p}$), we can replace eq. (A5) by
\begin{equation}
\theta \sim \theta_{0}+(\Omega-\Omega_{p})t. \\
\end{equation}
Using eq. (A8), full solutions of the linearized equations (A6 and A7) at the guiding center are written as
\begin{equation}
\xi = \sum_{m=1}^{n} \left[d_{m}{\rm cos}\hspace{0.1em}m(\theta-\vartheta_{m}(R))+e_{m}{\rm sin}\hspace{0.1em}m(\theta-\vartheta_{m}(R)) \right], \\
\end{equation}
\begin{equation}
\eta = \sum_{m=1}^{n} \left[ g_{m}{\rm sin}\hspace{0.1em}m(\theta-\vartheta_{m}(R))+f_{m}{\rm cos}\hspace{0.1em}m(\theta-\vartheta_{m}(R))\right],   \\
\end{equation}
where the amplitudes $d_{m}$, $e_{m}$, $g_{m}$, and $f_{m}$ are functions of Galactocentric distance as explained in the Appendix of Pi$\tilde{\rm{n}}$ol-Ferrer et al. (2012). Pi$\tilde{\rm{n}}$ol-Ferrer et al. (2012) and (2014) inserted the damping term $\varepsilon$ in the amplitudes to damp the corotation resonance. 

Finally, we describe simple relations between observed peculiar motions ($U$ and $V$) and the derived solutions ($\xi$ and $\eta$) as
\begin{equation}
\dot{\xi} = -U   \\
\end{equation}
and
\begin{equation}
\dot{\eta} = V.   \\
\end{equation}
Using equations (A11) and (A12), we can compare VLBI astrometry results with the analytic model. 
\clearpage

\section{The least-squares fit}
For comparisons between the analytic model explained in {\bf Appendix 1} and the VLBI astrometry results listed in table 5, we search the minimum value of the reduced chi-square ($\chi^{2}_{\nu}$) in analytic model parameters that are varied within appropriate ranges. The chi-square ($\chi^{2}$) can be simply written as
\begin{equation}
\chi^{2} = \sum_{i=1}^{n} \left[\frac{\Delta a_{i}}{\sigma_{i}}\right]^{2},   \\
\end{equation}
where $a_{i}$ is a measurement with an uncertainty $\sigma_{i}$, and $\Delta a_{i}$ is the residual between the measurement and an expected value. We use $U$ and $V$ with uncertainties from table 5 for the measurements, and the residuals are calculated using the measurements and the analytic model. 

Table 6 displays the analytic model parameters used for a spiral potential, while we assume a flat rotation model (i.e., $\Theta(R)$ = $\Theta_{0}$) and Galactic constants of ($R_{0}$, $\Theta_{0}$) = (8.33 kpc, 240.0 km s$^{-1}$) to describe a Galactic axisymmetric potential. There are seven parameters of which three parameters, the amplitude, pitch angle, and mode of the spiral potential, are fixed based on previous research. 

For the amplitude of the spiral potential, we refer to Grosb$\o$l et al. (2004) who showed the relation between pitch angles and amplitudes for spiral galaxies (see fig. 8 in Grosb$\o$l et al. 2004). The spiral amplitudes ranged between $\sim$ 0 and $\sim$ 50 $\%$ relative to disk amplitudes in Grosb$\o$l et al. (2004), and therefore we fix the amplitude between 10 and 50 $\%$ relative to the local amplitude with an increment of 10 $\%$. Note that the amplitude is related to surface density $\Sigma$ (see eq. 6.30 in Binney $\&$ Tremaine 2008) and we assume a local surface density ($\Sigma_{\odot}$) of 63.9 M$_{\odot}$ pc$^{-2}$ based on Mcmillan (2011).

For the pitch angle and mode of the spiral potential, we cite Vall\'ee (2013) who listed recent results of spiral arms in the Milky Way. The pitch angles between 2.3 (or 5.4) and 16.5 deg were listed with a mode of 2 or 4 in tables 1 and 2 of Vall\'ee (2013). Note that the pitch angle of 2.3 deg was determined based on only two sources associated with the Outer arm in Reid et al. (2009a), and recently the value was modified to 13.8 $\pm$ 3.3 deg based on six sources associated with the Outer arm in Reid et al. (2014b). Thus, we fix the pitch angle between 5 and 20 deg with an increment of 5 deg. Also, we fix the mode of 2 or 4.  

Aside from the fixed values, we may not have to fix the reference position ($R_{{\rm ref}}$) of the spiral potential based on VLBI astrometry results (i.e., $R_{{\rm ref}}$ = 10.4 kpc determined in section 4.1), since the VLBI astrometry results may not trace the bottom of the spiral potential. Therefore, we search it with an increment of 0.1 kpc in a range corresponding to phase between $-$$\pi$ and +$\pi$ radian of the spiral potential from $R_{{\rm ref}}$ = 10.4 kpc. Note that the search range of $R_{{\rm ref}}$ depends on not only the pitch angle, but also the mode of the spiral potential. For instance, in the case of i = 5.0$^{\circ}$ and m = 4, we search $R_{{\rm ref}}$ between 9.7 and 11.1 kpc, while we search $R_{{\rm ref}}$ between 5.8 and 18.2 kpc in the case of i = 20.0$^{\circ}$ and m = 2.

For the pattern speed ($\Omega_{p}$) of the spiral potential, we search it between 10 and 30 km s$^{-1}$ kpc$^{-1}$ with an increment of 0.5 km s$^{-1}$ kpc$^{-1}$ based on previous research (e.g., $\Omega_{p}$ = 11.5 $\pm$ 1.5 km s$^{-1}$ kpc$^{-1}$ in Gordon 1978; $\Omega_{p}$ = 30 km s$^{-1}$ kpc$^{-1}$ in Fern\'andez et al. 2001).

The other two parameters ($\lambda$ and $\varepsilon$) are not observables, and we search them in a relatively wide range. $\lambda$ and $\varepsilon$ are searched in the same range between 1 and 30 km s$^{-1}$ kpc$^{-1}$ with an increment of 0.5 km s$^{-1}$ kpc$^{-1}$. 

After the grid (discrete) search, we determine final values with errors using the formula provided by Bevington $\&$ Robinson (2002) as 
\begin{equation}
\chi^{2} = \frac{(a_{j}-a_{j}^{'})^{2}}{\sigma_{j}^{2}} + C,   \\
\end{equation}
where $a_{j}$ is a single parameter in the vicinity of the minimum (at $a_{j}^{'}$) of the $\chi^{2}$ distribution, $\sigma_{j}^{2}$ is a dispersion of the single parameter, and the constant $C$ is a function of the uncertainties $\sigma_{i}$ (as explained in eq. A13) and the other parameters $a_{k}$ for $k$ $\neq$ $j$. 

However, we note that the error obtained from eq. (A14) might be as a guide, since systematic error should be discussed in the spiral potential model.


\end{document}